\documentclass[11pt]{article}

\usepackage{fullpage}

\usepackage{natbib}
 \bibpunct[, ]{(}{)}{,}{a}{}{,}%

\usepackage{amsmath,amssymb,amsfonts,mathrsfs, mathtools, bm, bbm, dsfont, mathrsfs}
\usepackage{graphicx,xcolor}
\usepackage{ifthen}
\usepackage{multirow, multicol}
\usepackage{float,framed}
\usepackage{subeqnarray, array}
\usepackage{enumitem}
\usepackage{caption}
\usepackage{booktabs}
\usepackage{float}
\usepackage[caption=false]{subfig}
\usepackage{natbib}
\usepackage[font=small]{caption}
\usepackage[breaklinks, colorlinks, linkcolor=purple, anchorcolor=purple, citecolor=purple, urlcolor=purple]{hyperref}

\usepackage[ruled]{algorithm2e}
\usepackage{tikz}
\definecolor{salmon}{rgb}{1.0, 0.55, 0.41}

\definecolor{myblue}{RGB}{0, 128, 255}

\newcommand{\redline}{\raisebox{2pt}{\tikz{\draw[-,blue,solid,line width = 1.5pt](0,0) -- (6mm,0);}}}
\newcommand{\myblueline}{\raisebox{2pt}{\tikz{\draw[-,myblue,solid,line width = 1.5pt](0,0) -- (6mm,0);}}}
\newcommand{\blueline}{\raisebox{2pt}{\tikz{\draw[-,salmon,solid,line width = 1.5pt](0,0) -- (6mm,0);}}}

\newcommand{\beq}{\begin{equation}}
\newcommand{\eeq}{\end{equation}}
\newcommand{\beqa}{\begin{eqnarray}}
\newcommand{\eeqa}{\end{eqnarray}}
\newcommand{\beqan}{\begin{eqnarray*}}
\newcommand{\eeqan}{\end{eqnarray*}}

\newcommand\T{{\mathpalette\raiseT\intercal}}
\newcommand\raiseT[2]{\raisebox{0.25ex}{$#1#2$}
}

\newcommand{\Nset}{\mathds{N}}

\newcommand{\Rset}{\mathds{R}}

\newcommand{\Gcal}{{\cal G}}

\newcommand{\Ocal}{{\cal O}}
\newcommand{\Pcal}{{\cal P}}

\newcommand{\Wcal}{{\cal W}}

\newcommand{\bone}{\mathbf{1}}

\renewcommand{\v}[1]{{\bm{#1}}}

\renewcommand{\[}{\left[}
\renewcommand{\]}{\right]}

\renewcommand{\(}{\left(}
\renewcommand{\)}{\right)}

\newcounter{l1}
\newcounter{l2}
\newcounter{l3}
\newcommand{\bdotlist}{\begin{list}{$\bullet$}{}}
\newcommand{\bboxlist}{\begin{list}{$\Box$}{}}
\newcommand{\bbboxlist}{\begin{list}{\raisebox{.005in}{{\tiny
$\blacksquare$ \ \ }}}{}}
\newcommand{\bdashlist}{\begin{list}{$-$}{} }
\newcommand{\blist}{\begin{list}{}{} }
\newcommand{\barablist}{\begin{list}{\arabic{l1}}{\usecounter{l1}}}
\newcommand{\balphlist}{\begin{list}{(\alph{l2})}{\usecounter{l2}}}
\newcommand{\bAlphlist}{\begin{list}{\Alph{l2}.}{\usecounter{l2}}}
\newcommand{\bdiamlist}{\begin{list}{$\diamond$}{}}
\newcommand{\bromalist}{\begin{list}{(\roman{l3})}{\usecounter{l3}}}

\newtheorem{theorem}{Theorem}

\newtheorem{proposition}{Proposition}

\renewcommand{\bone}{\mathds{1}}
\newcommand{\CTS}{{\sf CTS}}
\newcommand{\TO}{{\sf TO}}
\newcommand{\NE}{{\sf NE}}
\newcommand{\SO}{{\sf SO}}

\newcommand{\utc}{{\sf UTC}}
\newcommand{\da}{{\sf DA}}

\usepackage{palatino}

\begin{document}


\title{\bf Coordinated Transaction Scheduling in Multi-Area Electricity Markets: Equilibrium and Learning}
\author{Mariola Ndrio \qquad Subhonmesh Bose \qquad Lang Tong \qquad Ye Guo} 


\maketitle

\begin{abstract}
Tie-line scheduling in multi-area power systems in the US largely proceeds through a market-based mechanism called Coordinated Transaction Scheduling (CTS). We analyze this market mechanism through a game-theoretic lens. Our analysis characterizes the effect of market liquidity, market participants' forecasts about inter-area price spreads, transactions fees and coupling of CTS markets with up-to-congestion virtual transactions. Using real data, we empirically verify that CTS bidders can employ simple learning algorithms to discover Nash equilibria that support the conclusions drawn from equilibrium analysis. 
\end{abstract}

\section{Introduction}
\label{sec:intro}

Different parts of an interconnected power grid are controlled and managed by different system operators (SOs). We call the geographical footprint within each SO's jurisdiction an area, and transmission lines that interconnect two different areas as tie-lines. Efficient scheduling of power flows over tie-lines is paramount to improve market efficiency and exploit geographically diverse renewable resources. Tie-lines are capable of supplying a significant portion of each area's electricity demand. For example, the New York ISO (NYISO) and ISO New England (ISO-NE) share nine tie-lines with approximately 1800 MW capacity, capable of supplying 12\% of New England's and 10\% of New York's demand as of 2009 (see \cite{pike}). 
Even though tie-lines are important assets, they have been historically under-utilized or scheduled in the counter-economic direction as \cite{pike} illustrate. The economic loss from inefficient tie-line schedules has been estimated at \$784 million between NYISO and ISO-NE in 2006-10 (see \cite{pike}), the burden of which has been ultimately borne by end-use customers. What causes such inefficiencies? There are a number of factors including the inherent uncertainty about power requirements when tie-lines are scheduled prior to delivery time points, the lack of coordination among SOs, ad hoc use of designated trading locations and transaction fees.

Conceptually, power flows over tie-lines should be determined through a joint economic dispatch problem geared towards maximizing the efficiency of the interconnected power grid as a whole. However, historical and legal reasons render such an aggregation of market information from different areas at a central location untenable. Naturally, a considerable effort has been made to solve the joint dispatch problem in a distributed fashion, focusing on primal (e.g., \cite{biskas,zhao}) and dual decomposition methods (see \cite{conejo,Chen,Kim}). In such methods, SOs exchange information among themselves to compute the optimal tie-line schedule. This theoretical coordination mechanism, referred to as Tie Optimization (TO) by \cite{pike}, proved challenging to implement in practice. It was perceived as requiring the SOs to trade directly with each other, violating their financial neutrality, in lieu of the earlier market-based, albeit inefficient, process for scheduling tie-line flows. Instead, many SOs adopted variants of Coordinated Transaction Scheduling (CTS), e.g., see \cite{cts1,cts2}, that sought to blend the earlier market-based tie-line scheduling with the theoretically optimal TO, after receiving approval from FERC. CTS is a market mechanism in which external market participants submit bids and offers to import or export from one area to the other. CTS market design is predicated on the simple premise that arbitrage opportunity will attract more participants, whose profit motivation will ultimately shrink that opportunity, pushing the schedule closer to the theoretically optimum. CTS has certainly improved tie-line scheduling as per \cite{isone,nyiso}, but significant inefficiencies remain. Motivated by these inefficiencies, we present a theoretical model to analyze CTS and investigate the repercussions of strategic behavior on overall market performance. We provide palpable insights on the consequences of an illiquid market, errors in SOs' price forecasts and transaction fees on market efficiency, all of which have been named in \cite{nyiso} as crucial factors affecting CTS market efficiency.\footnote{We remark that the use of designated trading locations for CTS transactions results in the so-called `loop flow' problem (see \cite{ilic1}) that negatively impacts CTS market performance. We refer the reader to \cite{guo} for mechanisms to tackle this problem.}

We introduce the mechanics of CTS in Section \ref{sec:mechanisms}. Then, we model CTS as a game among arbitrage bidders who compete through scalar-parameterized transport offers in Section \ref{sec:gameCTS}. To the best of our knowledge, this is the first work that provides a concrete mathematical formulation to model CTS as a game. Our formulation is inspired by supply function competition models considered by \cite{johari,Xu}. While general supply function competition models have a long history in the analysis of electricity markets (see \cite{Rudkevich}), our scalar-parameterized supply function competition model makes it particularly amenable to theoretical analysis that may be of independent interest.

We establish the existence of Nash equilibria for this game and study the impact of various factors on the nature of said equilibria in Sections \ref{sec:affine}-\ref{sec:conjectures} to offer insights into the CTS market. First, we show that when transaction costs (levied on a per-megawatt hour basis on bidders) are absent, then a highly liquid CTS market is efficient. Market efficiency degrades with liquidity shortfall, exhibiting bounded efficiency loss for intermediate liquidity and unbounded losses in low liquidity regimes. Second, with transaction costs, CTS fails to eradicate the price spread between adjacent markets even with a liquid market, implying that such costs undercut the vision behind the market design. Third, we show that SOs' estimate of the price spread plays a central role in the efficiency of CTS markets in that bidders have limited ability to correct the effects of SOs' forecast errors. Fourth, portfolios of virtual transactions such as up-to-congestion (UTC) bids held by CTS bidders can impact CTS market outcomes, revealing the dependency of efficiency of these inter-area markets on other energy markets. Our equilibrium analysis reveals how the strategic incentives in CTS markets are oriented but does not illustrate if bidders can learn equilibrium behavior through repeated participation in these markets. We simulate repeated play using historical data from the NYISO--ISO-NE market. In particular, we allow bidders to update their bids through a well-known upper confidence bound (UCB) algorithm that has been well studied in the reinforcement learning literature. Our simulations confirm that our conclusions from equilibrium analysis continue to hold in a statistical sense in our numerical experiments. 
All proofs are relegated to the Appendix.

\section{The CTS Mechanism}
\label{sec:mechanisms}

\begin{figure}[!h]
	\centering
	\includegraphics[width=0.65\textwidth]{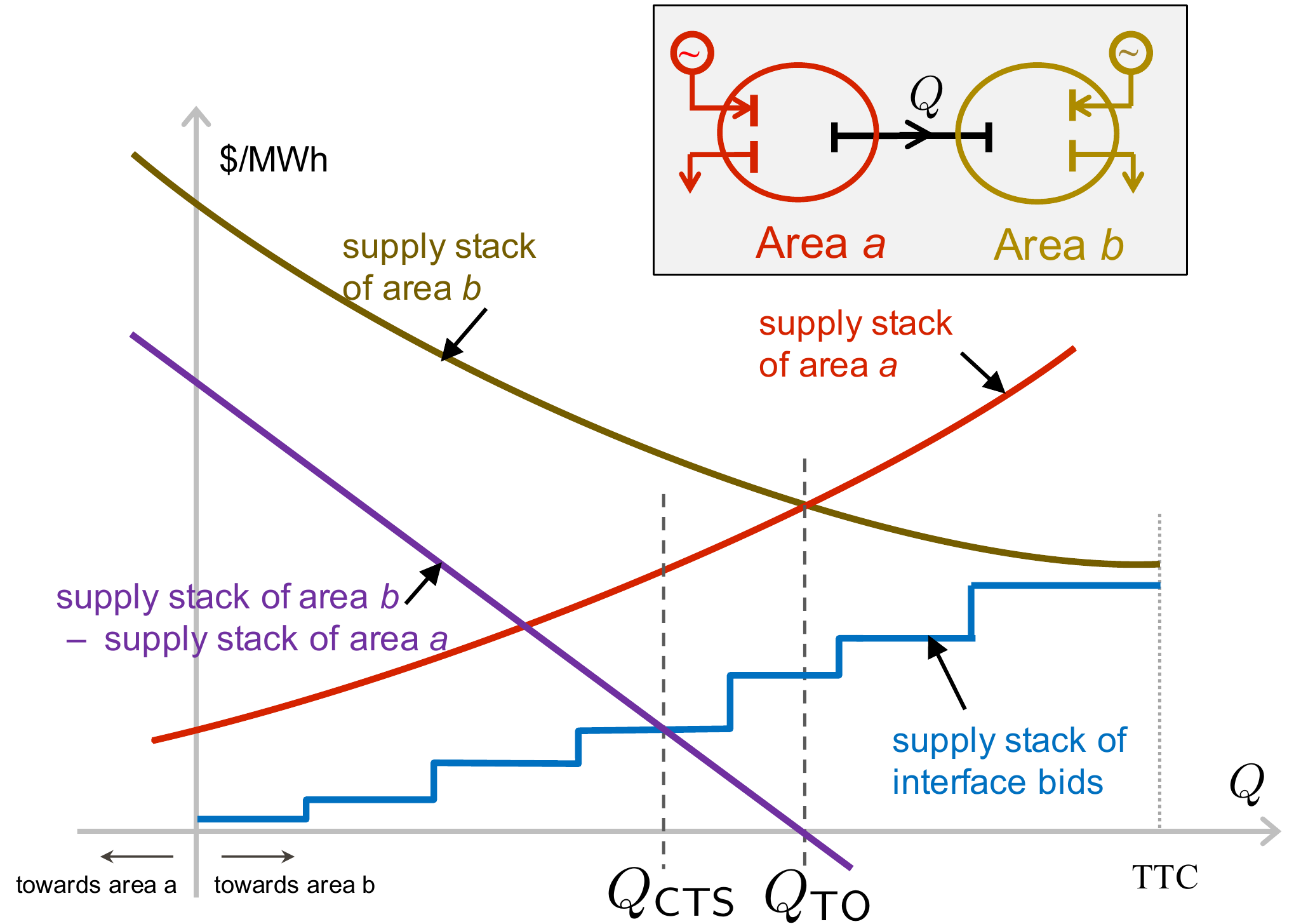}
	\caption{Illustration of the TO and CTS scheduling mechanisms.}
	\label{mechanisms}
\end{figure}

CTS is a real-time, market-based mechanism for tie-line scheduling that replaced an earlier market-based structure in an effort to streamline the bidding and scheduling process. Among the important changes, CTS unified the bid submission and clearing process between neighboring SOs, reduced the tie-line schedule duration from one hour to 15-minute intervals, and decreased time delays among bidding, scheduling, and power delivery. To illustrate the mechanics and economic rationale of CTS, consider two power systems connected via a common interface, with the interface power flow denoted by $Q$ as shown in Figure \ref{mechanisms}. Assume the SOs want to determine the tie-line schedule for an upcoming interval $[t, t+15]$. Then, at $t-15$ each SO computes a supply stack by solving an area-wise parametric economic dispatch with varying values of $Q$.
An example of supply stack is shown in Figure \ref{mechanisms}. In this example, at $Q=0$, area $b$ operates at higher costs than area $a$. Hence, area $a$ is the exporting region and area $b$ the importing, i.e., the direction of $Q$ is from $a$ to $b$. The supply stack of area $a$ represents the expected incremental dispatch cost of delivering power at a designated trading location--a node in the network in geographical proximity to area $b$ but not necessarily associated with the tie-lines. Similarly, the stack of area $b$ represents the expected decremental dispatch cost of reduced supply, shown in descending order. In other words, these supply stacks represent how the locational marginal price (LMP) at each SO's designated trading location, varies with $Q$.
At the level where supply stacks intersect, the tie-line schedule minimizes the aggregate dispatch costs across the two areas. This schedule, denoted by $Q_\TO$, corresponds to the outcome of a theoretical scheme referred to as tie optimization (TO).
While CTS remains our focus in this paper, TO serves as our theoretical benchmark to compare CTS against. 
Contrary to TO, CTS relies on virtual traders whose offers/bids are utilized together with the supply stacks to arrive at the tie-line schedule, as we describe next.

A CTS participant is a virtual bidder that can offer to transport power across areas without physically consuming or producing it. They only participate in the tie-line scheduling process, bearing no obligation for physical power delivery; the transaction is purely financial. In particular, CTS participants submit interface bids that consist of three elements: the minimum price difference the bidder is willing to accept, the maximum quantity to be transferred, and the direction of trade, i.e., the exporting and importing area. All the bids indicating a direction from $a$ to $b$ are stacked from lowest to highest price, to create their own interface supply stack as shown in Figure \ref{mechanisms}. Bids that indicate direction from $b$ to $a$ are rejected at the outset since they would widen the SO-predicted price spread.
The price spread curve is derived by subtracting the supply stack of area $a$ from that of area $b$. The CTS schedule, denoted by $Q_\CTS$, is set at the intersection of the interface supply stack and the price spread.\footnote{The intersection of the supply stacks can occur to the right of the total transfer capability (TTC) of the interface. In such cases, $Q_\CTS$ is equal to TTC, preventing price convergence. However, according to \cite{pike}, the primary interface between NYISO--ISO-NE was congested 0.3\% and 1.2\% of the hours eastbound and westbound, respectively, in 2009. In this work, we focus on the factors that cause price separation under CTS, other than TTC.} An interface bid is accepted if its offer price is less than the price spread at the tie-line schedule. Therefore, all interface bids to the left of the CTS schedule are accepted; all bids to the right are not.

CTS bids can be submitted up to $t-75$, are cleared at $t-15$ and are settled at the ex-post LMPs calculated for the time period $[t, t+15]$. Hence, there is approximately a 30-minute latency time for the SOs and 90 minutes for CTS participants. This latency problem exposes participants to financial risk since there is uncertainty at which LMPs CTS bids will settle. LMPs are highly volatile (see Figure II-7 in \cite{pike}) and CTS bids that appeared financially favorable at $t-15$ may become unfavorable at $t+15$. Such risks impact bidding behavior and possibly CTS market efficiency. It may not be possible to eliminate latency completely as tie-line schedules are typically decided with a lead time to power delivery.
	
In the sequel, we present a theoretical abstraction that models crucial features of CTS but also permits rigorous mathematical analysis. This model allows us to identify several factors that negatively impact CTS markets. We take the viewpoint that if our stylized CTS model reveals design inefficiencies, practical market considerations will likely add to such inefficiencies, making the market perform even more poorly than our analysis suggests.

\section{Modeling the CTS Market as a Game}
\label{sec:gameCTS}
The first question we answer is whether the incentives of CTS bidders are aligned with those of the SOs and CTS design.
To reveal the impacts of bidding behavior on CTS, we model CTS as a game among virtual bidders who compete to transport power over the tie-lines against an elastic inter-area price spread that varies with $Q$.  
For areas $a$ and $b$, denote by $\Pcal_a(Q)$ and $\Pcal_b(Q)$ the LMPs at CTS trading locations, respectively. Without loss of generality, let area $a$ export and area $b$ import power, and define
\begin{equation}
	\Pcal(Q) := \Pcal_b (Q) - \Pcal_a(Q)\label{eq:priceSpread}
\end{equation}
as the price spread between the areas.

\textbf{Assumption 1.} $\Pcal : \Rset \rightarrow \Rset$ is differentiable, concave, invertible and strictly decreasing in $Q\geq 0$ with $\Pcal(0)>0$.

Concavity, differentiability and decreasing nature of $\Pcal$ are standard assumptions in prior literature on supply function and Cournot competition models, e.g., see \cite{Green,baldick,Rudkevich,Meyer}. These assumptions facilitate the game-theoretic analysis.

\begin{figure}[H]
	\centering
	\includegraphics[width=0.4\textwidth]{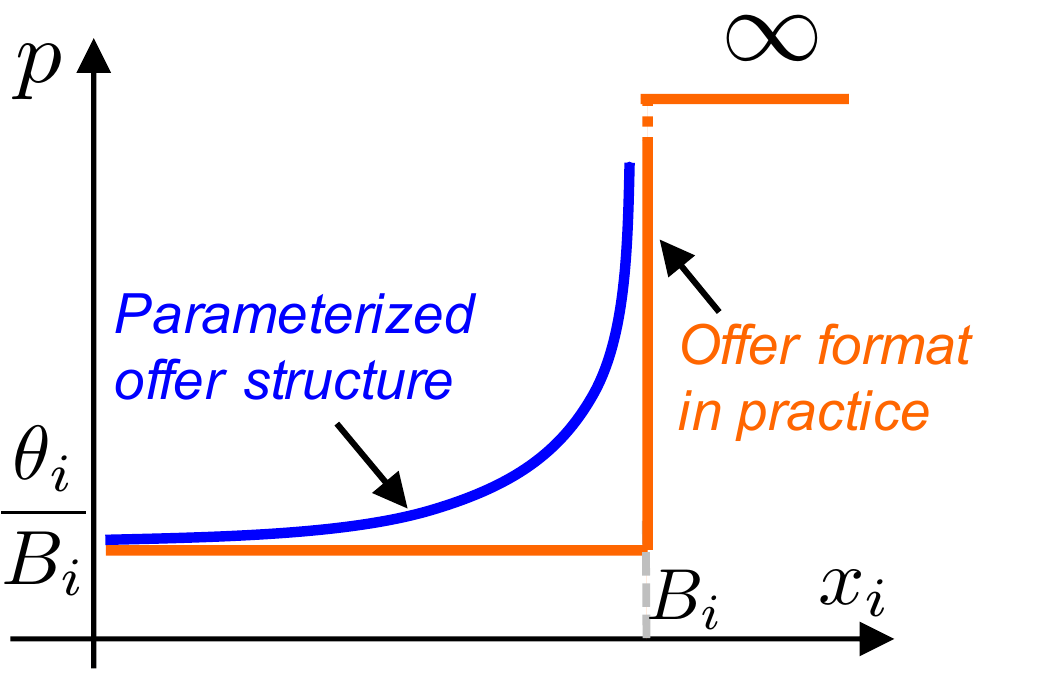}
	\caption{Parameterized interface bid of CTS market participant.}
	\label{fig:supplyFunction}
\end{figure}

Consider $N$ virtual bidders in the CTS market. Let bidder $i$ provide two parameters $\theta_i, B_i$ to the SOs with the understanding that she is willing to transport up to 
\begin{align}
	x_i(p) := B_i - \frac{\theta_i}{p}, \quad \theta_i \geq 0
	\label{eq:supplyOffer}
\end{align}
amount of power from area $a$ to $b$ at a price spread of $p >0$. Our transport offer is inspired by supply function competition models studied in \cite{johari,Xu,Ndrio2}.
Figure \ref{fig:supplyFunction} reveals how the parameters $\theta_i, B_i$ affect the shape of the transport offer. Bidder $i$ is willing to transport a maximum quantity of $B_i$, but at a minimum price spread of $\theta_i/B_i$. The required price difference increases with the power transport and grows unbounded as the latter approaches $B_i$. In effect, transporting power above $B_i$ requires an infinite price difference. Therefore, bidder $i$ expresses her total budget or her liquidity in $B_i$. In what follows, we assume that the bidder acts strategically in $\theta_i$, given $B_i$ that models her budget constraints. Limiting the quantity to less than equal to $B_i$ limits the financial risk exposure of a market participant.\footnote{We remark that the transport offer considered in \eqref{eq:supplyOffer} allows $x_i$ to be negative. However, such outcomes do not arise at an equilibrium.}

Using the discrete blocks of price-quantity pairs, the characterization of equilibria may be analytically intractable, even in the absence of network constraints (see \cite{Anderson,Philpott}). The advantage of ``hockey-stick'' shaped transport offers in \eqref{eq:supplyOffer} is that they act as a smooth approximation to the interface bid used in practice, greatly facilitating the analysis.
Using the family of transport offers in \eqref{eq:supplyOffer}, the bidders participate in a capacitated scalar-parameterized supply function competition against the elastic demand in \eqref{eq:priceSpread}. Starting from the seminal work of Klemperer and Meyer \cite{Meyer}, supply function competition has been extensively studied over three decades with application to electricity market analysis (see \cite{Rudkevich}). Scalar-parameterized supply functions are more conducive to analysis and yield attractive results in terms of efficiency loss as \cite{johari, Xu,Lin} have shown. Moreover, said offer structures are more expressive than pure price (Bertrand) or quantity (Cournot) competition models that do not adequately capture a CTS interface bid. Our model goes one step beyond and considers these transport offers with an elastic demand that naturally applies to CTS markets. 

Given the liquidities $\v{B} = (B_1, \ldots, B_N)$, the choice of bids $\v{\theta} = (\theta_1, \ldots, \theta_N)$ from the CTS bidders describes their willingness to transport power across the interface according to \eqref{eq:supplyOffer}. The SOs calculate $\v{x}^\star :=(x_1^\star, \ldots, x_N^\star)$ as the allocations of the tie-line flow to the participants by solving

\begin{equation}
	\v{x}^\star(\v{\theta};\v{B}) \in \ \underset{\v{x} \leq \v{B}}{\text{argmax}} \int_{0}^{\bone^\T \ \v{x}}\Pcal(z)dz - \sum_{i=1}^{N} \int_{0}^{x_{i}}\dfrac{\theta_{i}}{B_{i} - s} ds, 
	\label{eq:marketProblem}
\end{equation}
where $\bone$ denotes a vector of ones of appropriate size. Notice that the transport offer enters the SOs' problem as the `bid-in cost' of each CTS bidder to transport quantity $x_i$. With this interpretation, the SOs' flow allocation problem in \eqref{eq:marketProblem} seeks to maximize the social welfare of an economy that is composed of the wholesale markets in areas $a$ and $b$ together with the CTS bidders (see \cite{guo} for a similar interpretation of the CTS market objective). 

The CTS schedule occurs where the offer stack for inter-area power transport from CTS market participants intersects the SOs' price spread function. Formally,
\begin{equation}
	\Pcal(Q_{\CTS}) = \dfrac{\bone^\T \ \v{\theta}}{\bone^\T \ \v{B} - Q_{\CTS}}, \ \textrm{for } \ \bone^\T \ \v{\theta} > 0. \label{eq:Qcts}
\end{equation}
Denote the solution of \eqref{eq:Qcts} by $Q(\v{\theta}; \v{B})$. Then, the market clearing price is given by
\begin{equation}
	p(\v{\theta};\v{B}) = \Pcal(Q_\CTS (\v{\theta}; \v{B}) ).
	\label{eq:market price}
\end{equation} 
We define a useful benchmark: the maximum inter-area demand or $Q_{\TO}$. This schedule corresponds to the quantity for which the inter-area price spread vanishes or formally
\begin{equation}
	Q_{\TO} \in \underset{Q \geq 0}{\text{argmax}} \ \Wcal(Q):=\int_{0}^{Q} \Pcal(z)dz. \label{eq:TO}
\end{equation}
At $Q_{\TO}$ there is no more opportunity for arbitrage as $\mathcal{P}(Q_{\TO}) = 0$. The CTS flow allocation to every participant is given by
\begin{equation}
	x_i^\star(\v{\theta}; \v{B}) = B_i - \dfrac{\theta_i}{ p(\v{\theta};\v{B})  },  \ \textrm{for  } \bone^\T \ \v{\theta} > 0.
\end{equation}
When $\bone^\T \ \v{\theta} = 0$, from \eqref{eq:marketProblem} it follows that $Q_{\CTS} = \min\{ \bone^\T \ \v{B} , Q_{\TO} \}$. When $\bone^\T \ \v{B} < Q_{\TO}$, $x_i^\star(\v{0}; \v{B}) = B_i$ irrespective of $p$. On the other hand, if $\bone^\T \ \v{B} \geq Q_{\TO}$, then any feasible solution of \eqref{eq:marketProblem} is optimal. In this case, we specify $x_i^\star$ as the allocation of $Q_{\TO}$ proportional to each participant's budget, i.e., $x_i^\star (\v{0}; \v{B}) = (B_i / \bone^\T \ \v{B}) Q_{\TO}$. With these additional conventions, $x_i^\star$ is well-defined for any $\v{\theta}$ and $\v{B}$.

While virtual bidders do not incur any costs to physically transport power, many pairs of SOs levy transaction fees on a per-MWh basis, e.g., in CTS between NYISO and PJM, NYISO charges physical exports to PJM at a rate ranging from \$4-\$8 per MWh, while PJM charges physical imports and exports rates that average less than \$3 per MWh.
See \cite{nyiso} for details.  
For a willingness to transport $x_i$ MW of power from area $a$ to $b$, assume that transaction cost equals $c\cdot x_i$, where $c$ is measured in \$/MWh. Then, each bidder's payoff equals the total revenue garnered less the transaction costs, formally given in
\begin{align}
	\pi_{i}(\theta_i, \v{\theta}_{-i}) 
	&=\Pcal\left( Q_\CTS(\v{\theta}; \v{B}) \right) x_{i}^{\star}(\v{\theta};\v{B}) - cx_{i}^{\star}(\v{\theta}; \v{B})\notag \\
	& = \Pcal\left( Q_\CTS(\v{\theta}; \v{B}) \right)B_i - \theta_i - cx_{i}^{\star}(\v{\theta}; \v{B}),
	\label{eq:payoff.general}
\end{align} 
where $\v{\theta}_{-i}$ denotes a vector with all but the $i$th component of $\pmb{\theta}$.

With this discussion in mind, we now proceed to define the CTS game. The set of players consists of $N$ CTS participants. When players incur costs $c\geq 0$, any player bidding $\theta_i / B_i \leq c$ would incur a loss. Hence, we restrict each player's actions to satisfy $\theta_i \geq c B_i$. Define $\Gcal(\v{B},c)$ as the CTS game among $N$ virtual bidders who bid $\theta_i$, given $\v{B}$, and receive a payoff described by \eqref{eq:payoff.general}. Bidders selfishly seek to maximize their own payoffs, given their liquidities. 
A bid profile $\v{\theta}^\NE$ constitutes a \emph{Nash equilibrium} of $\Gcal(\v{B},c)$,  if
$$ \pi_{i}\left(\theta_i^\NE, \v{\theta}_{-i}^\NE\right) \geq \pi_{i}\left(\theta_i, \v{\theta}_{-i}^\NE\right)$$
for all $\theta_i \geq cB_i$. That is, no player has an incentive for a unilateral deviation from the equilibrium offer. Notice that implicit in the calculation of Nash equilibria is the assumption that players know $\v{B}$, which is unrealistic in practice. However, in Section \ref{sec:simulation} we empirically show that an equilibrium can be learned through repeated play even under settings when the perfect information assumption is relaxed. This ``learnability" lends credence to the conclusions drawn from our equilibrium analysis.
With this discussion in mind, we establish the existence of such an equilibrium profile in our first result. 
\begin{theorem}[Existence of Nash Equilibrium]
	\label{th:general}
	Let Assumption 1 hold. Then, the CTS game $\Gcal(\v{B},c)$ admits a Nash equilibrium if $\Pcal$ satisfies 
	\begin{equation}\Pcal ''(Q) (\bone^\T \ \v{B} - Q) \geq 2 \Pcal '(Q). \label{eq:existence}\end{equation}
\end{theorem}	
Our proof relies on Rosen's result in \cite{rosen} after we establish that $\Gcal(\v{B},c)$ is a concave game. Existence of an equilibrium requires the additional condition on $\Pcal$ given by \eqref{eq:existence} that is satisfied by many commonly used demand function families including affine models. Given that oligopolies with scalar-parameterized offers have been analyzed under inelastic demands (e.g. see \cite{johari, Xu, Lin}), Theorem \ref{th:general} is of independent interest as it establishes existence of Nash equilibria in settings with elastic demands. To explicitly characterize the Nash equilibrium, we restrict our attention to affine price spreads
\begin{equation}
	\Pcal(Q) := \alpha - \beta Q
	\label{eq:affine}
\end{equation}
with $\alpha, \beta > 0$. Therefore, from Theorem \ref{th:general} we conclude that an equilibrium always exists for $\Gcal(\v{B}, c, \alpha, \beta)$. The price spread can be shown to be affine in $Q$, when each area is represented as a copperplate power system, having a generator with quadratic generation costs and a fixed demand. This follows from properties of multiparametric quadratic programs in \cite[Theorem 7.6]{borelli}. To further justify our modeling choice, 
we perform a linear regression of New England's LMP at the CTS location (Roseton) as $ \Pcal_{\text{NE}} = w_1 \Pcal_{\text{NY}} + w_2 Q + w_3$, where $\Pcal_{\text{NY}}$ is the LMP at New York's CTS location. We obtain $w_1 \approx 1.0$ with an adjusted $R^2$ coefficient of 0.95, revealing an affine dependency of 
$\Pcal_{\text{NY}} - \Pcal_{\text{NE}}$ in $Q$. We obtain similar results when $\Pcal_{\text{NY}}$ is the dependent variable and $\Pcal_{\text{NE}}, Q$ are used as predictors \footnote{Regression analysis and data are available online: https://github.com/Mariola-Nd/CTS.git}.
However, expecting a perfectly linear relationship between $Q$ and $\Pcal$ over a period of, say a year, is a tall order. Spreads are typically noisy and depend on multiple factors such as renewable generation (see \cite{Woo}), fuel prices (see \cite{Carmona}), seasonality (see \cite{Karakatsani, Fanelli}), etc. However, given the rough affine dependency of the price spread from our data analysis, we restrict attention to affine $\Pcal$ for the rest of the paper. We emphasize that our goal in this work is not to perfectly forecast inter-area price spread functions, but to reveal market design flaws. When our game-theoretic analysis with affine $\Pcal$ reveals inefficient outcomes, practical considerations will possibly only exacerbate it. An affine $\Pcal$ suffices to reveal such inefficiencies.

\section{Impact of Liquidity in CTS Markets}
\label{sec:affine}
\begin{figure*}
	\centering
	\subfloat[]{\includegraphics[width=0.32\linewidth]{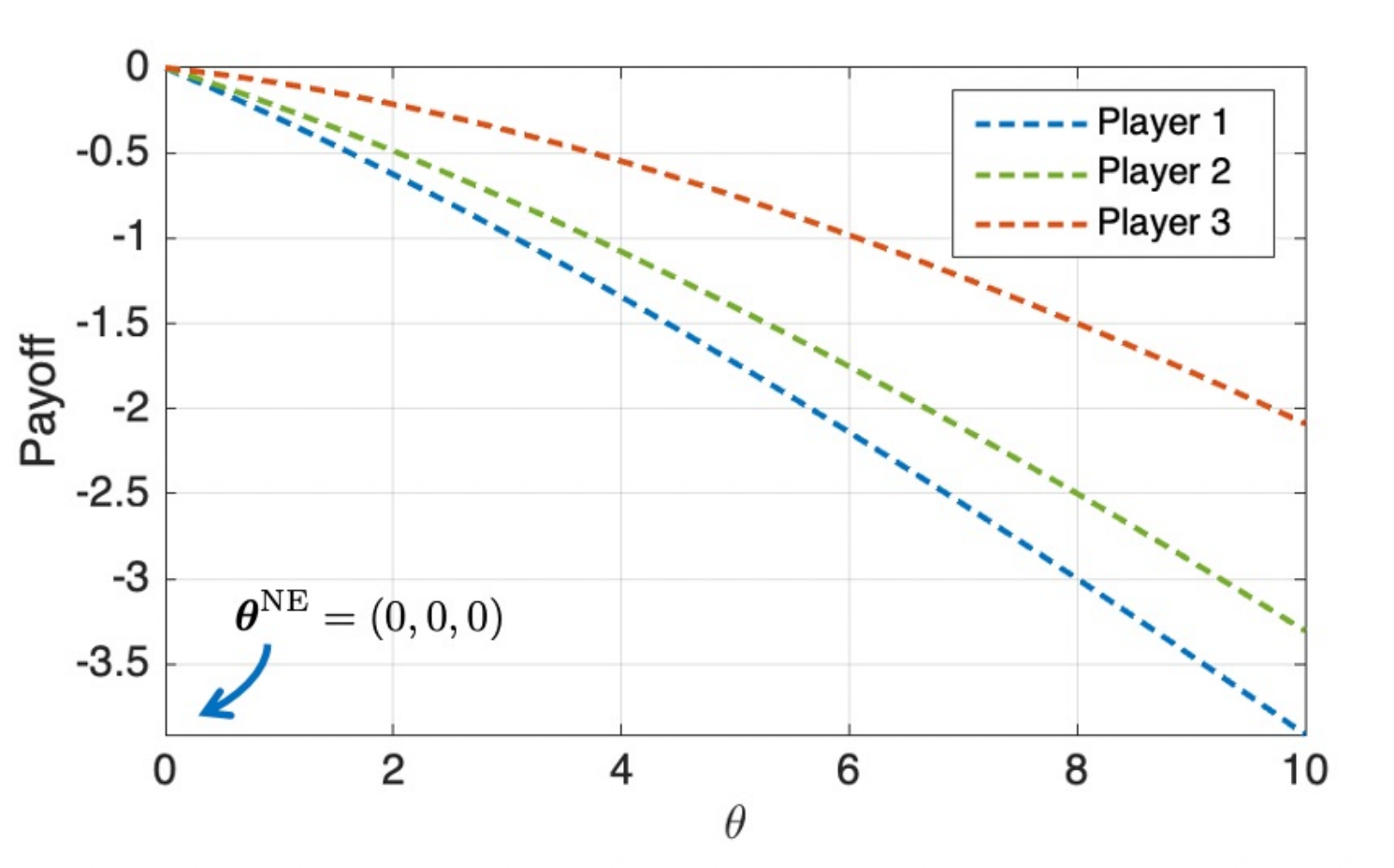} \label{fig:high}}
	\subfloat[]{\includegraphics[width=0.32\linewidth]{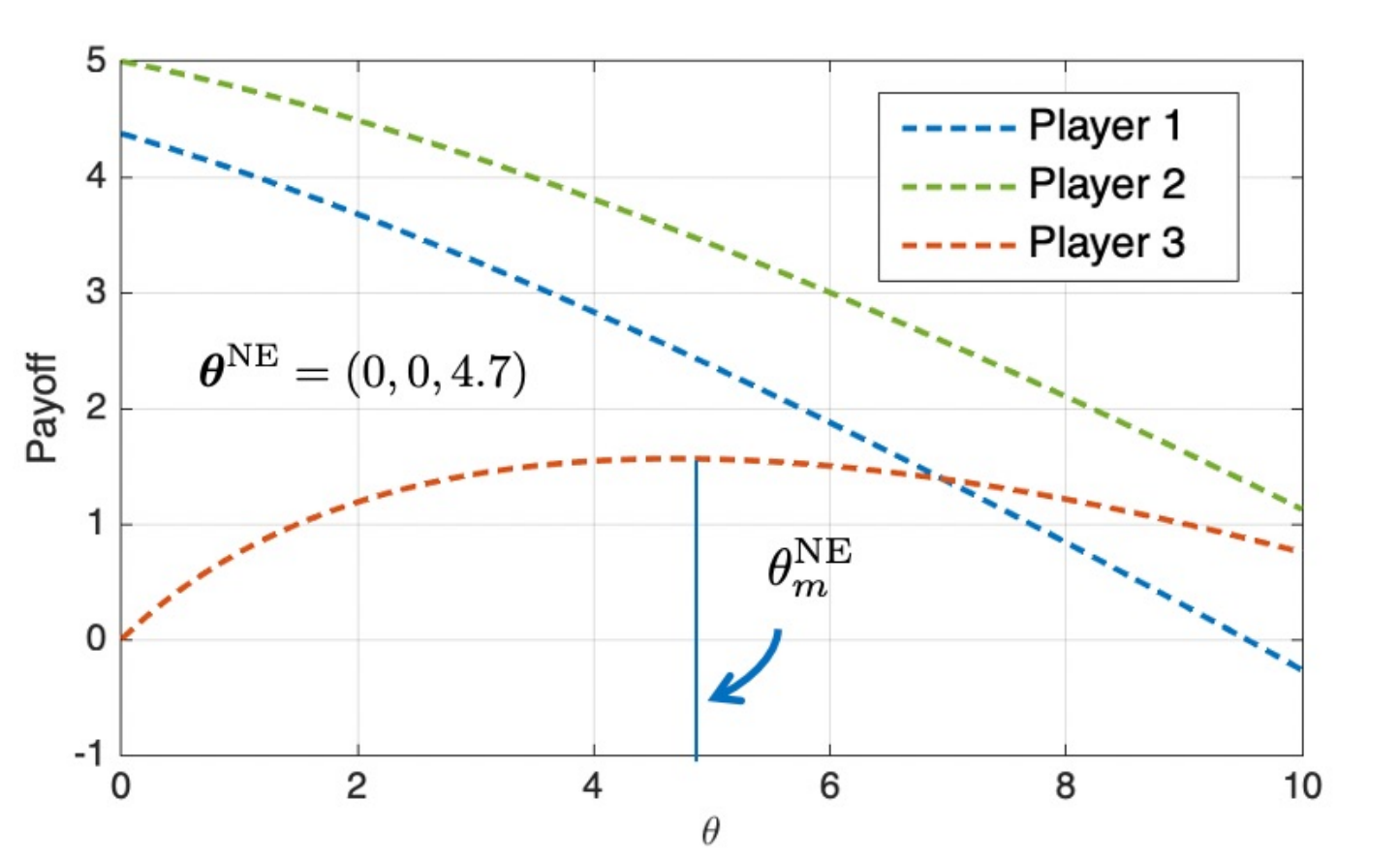} \label{fig:inter}}
	\subfloat[]{\includegraphics[width=0.32\linewidth]{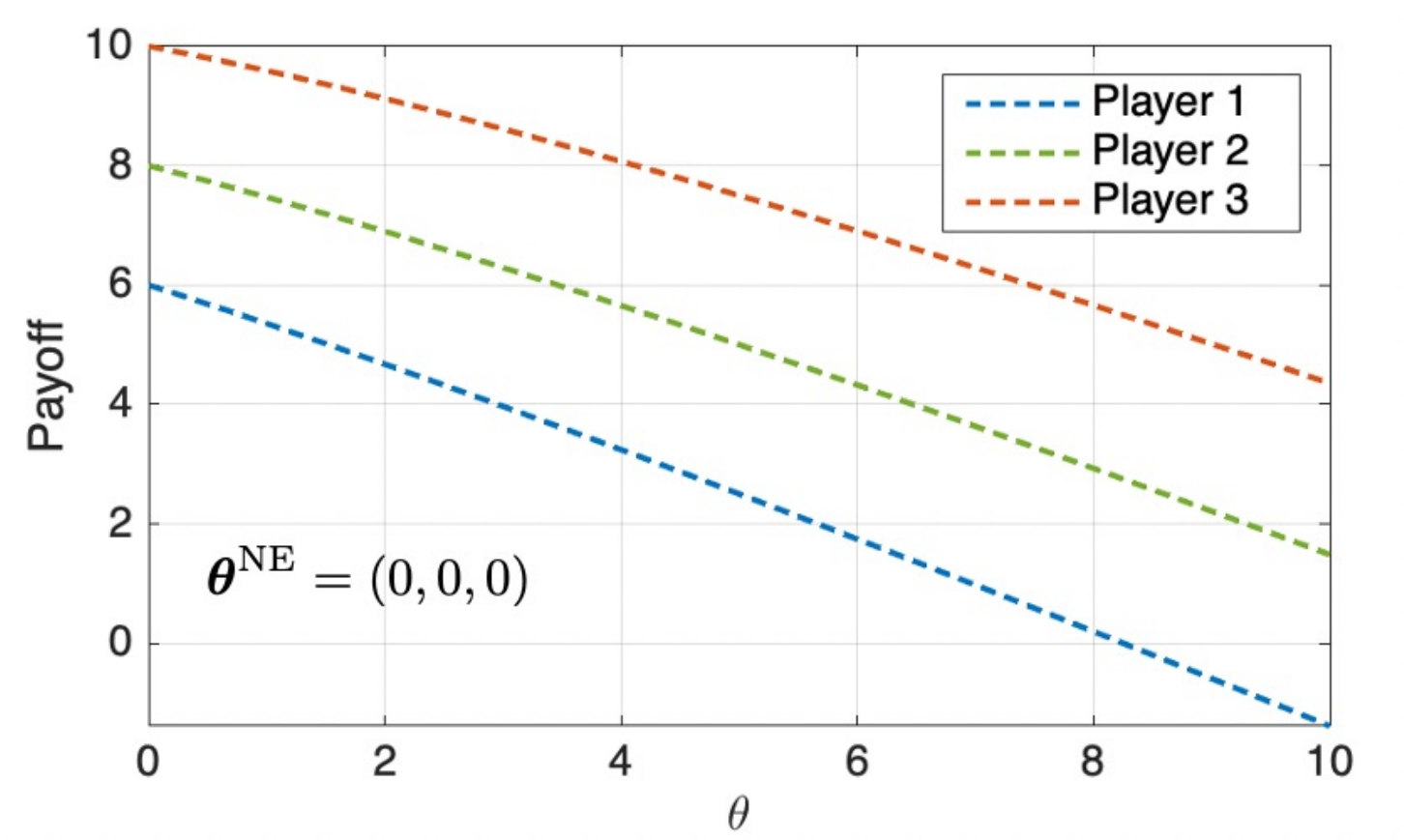} \label{fig:low}}
	\caption{Plots (a), (b) and (c) show payoffs of a 3-player CTS game $\Gcal(\v{B}, 0, \alpha, \beta)$ in the high, intermediate and low liquidity regimes, respectively. The liquidities satisfy $B_1 < B_2 < B_3$.}
	\label{fig:payoffs}
\end{figure*}
Our first goal is to investigate the impacts of liquidity on the CTS scheduling efficiency. To isolate the effects of liquidity, neglect transaction fees and set $c \approx 0$. 
We define the efficiency of CTS as the ratio
$$ \eta_{\CTS}(\v{B}) := \dfrac{\Wcal\left(Q_{\CTS}({\v{\theta}^{\NE}}, \v{B})\right)}{\Wcal\left(  Q_{\TO}\right)},$$ 
where recall that $\Wcal$ measures the aggregate welfare of the wholesale markets in the two areas attained at a particular interface schedule. TO seeks to maximize this welfare with $Q_{\TO} = \alpha/\beta$, while the outcome of CTS arises from the strategic interaction of the market participants. 

Our next result characterizes the equilibrium and provides key insights into the behavior of $\eta_\CTS \leq 1$ in different liquidity regimes. 
\begin{proposition}
	\label{prop:Nash}
	Consider the CTS game $\mathcal{G}(\v{B},0, \alpha, \beta)$, where $B_m$ is the unique maximal budget in $\left\{B_1, \ldots, B_N\right\}$. Then, $\mathcal{G}(\v{B},0, \alpha, \beta)$ admits a unique Nash equilibrium $\v{\theta}^{\NE}$ given by
	\begin{equation}
	\theta_m^{\NE} = \begin{cases} \dfrac{1}{4\beta}\left(\beta^2 B_m - \Pcal^2(\bone^\T \ \v{B})\right), & \text{if }  \left|  \bone^\T \ \v{B} -{\alpha}/{\beta} \right| < B_m, \\
	0, & \text{otherwise},
	\end{cases}
	\end{equation}
	and $\theta_i^{\NE} = 0$ for $i \neq m$. Furthermore, we have
	\begin{equation}
	\eta_{\CTS}(\v{B}) \begin{cases}
	= 1, & \text{if }  \bone^\T \ \v{B} - \alpha / \beta \geq B_m, \\
	\geq \dfrac{3}{4}, & \text{if }  \left| \bone^\T \ \v{B} - {\alpha}/{\beta} \right| < B_m,\\
	= 2z - z^2, & \text{otherwise}
	\end{cases},
	\end{equation}
	where $z:= \frac{\beta}{\alpha}\bone^\T \ \v{B}$.
\end{proposition}
The result highlights that allocation and efficiency vary widely with liquidity and the player with the maximal liquidity plays a rather central role in determining the outcome of the CTS market. To offer more insights, distinguish three different liquidity regimes. Identify the liquidity as high when $\bone^\T \ \v{B} - \alpha/\beta \geq B_m$, where the aggregate liquidity of all players but $m$ is sufficient to cover the efficient schedule $Q_\TO = \alpha/\beta$. The intermediate liquidity occurs where the aggregate liquidity is different from $Q_\TO$ by at most the liquidity of player $m$, i.e., 
$ \left| \bone^\T \ \v{B} - {\alpha}/{\beta} \right| < B_m$. Finally, the low liquidity regime is where $\bone^\T \ \v{B} + B_m < Q_\TO$. The outcome and the efficiency differ substantially across these regimes.
Using the equilibrium profile, it is easy to see that the flow allocation is given by 

\begin{align*}
x_m^\star({\v{\theta}^{\NE}}; \v{B}) 
&= \begin{cases} \frac{1}{2}( {\alpha}/{\beta} - \bone^\T \ \v{B}_{-m}), &\text{if } \left| \bone^\T \ \v{B} - {\alpha}/{\beta} \right| < B_m, \\
B_m, &\text{otherwise},
\end{cases}, 
\\
x_i^\star({\v{\theta}^{\NE}}; \v{B}) 
&= B_i,  \quad i \neq m,
\end{align*} 
where $\v{B}_{-m}$ denotes the vector of liquidities of all players, except $m$. Thus, all but player $m$ offer their maximum liquidity at equilibrium. These players benefit from being inframarginal, exploiting the bid of the marginal player $m$. This behavior is reminiscent of the so-called `free-rider problem' (see \cite{fudenberg}).
When the liquidity is too high or too low, player $m$ does not have enough market power and does not benefit from bidding nonzero $\theta_m$, implying that she does not withhold from her maximal budget $B_m$ in her transport offer. In the intermediate liquidity case, player $m$ enjoys market power and her flow allocation can be shown to be the Cournot best response to this residual price spread $\Pcal(Q - \bone^\T \ \v{B}_{-m})$. 

The tie-line schedule at the equilibrium of $\Gcal(\v{B},0, \alpha, \beta)$ is 
\begin{align*}
Q_{\CTS} &= \begin{cases}
Q_{\TO}, &\text{if } \bone^\T \ \v{B} - B_m \geq \alpha / \beta, \\
\frac{1}{2} \left( Q_{\TO} + \bone^\T \ \v{B}_{-m}\right), & \text{if } \left| \bone^\T \ \v{B} - {\alpha}/{\beta} \right| < B_m, \\
\bone^\T \ \v{B}, &\text{otherwise}.
\end{cases}
\end{align*}
When liquidity is high, $Q_\CTS$ coincides with $Q_\TO$, implying that CTS yields the SOs' intended outcome. In other words, perfect competition arises as a result of strategic incentives. In the intermediate liquidity regime, CTS suffers welfare loss due to strategic interaction. The loss, however, is bounded; strategic behavior cannot cripple the welfare under perfect competition by more than 25\%. When the liquidity is low, the lower bound on $\eta_\CTS$ can be arbitrarily small. However, in this case, lack of efficiency is not due to strategic interactions but rather due to the lack of market liquidity.

To offer further insights in the previous discussion, consider an example of a CTS game with three players. The players' payoffs are shown in Figure \ref{fig:payoffs} for each liquidity regime. When liquidity is high, all players garner zero payoffs by bidding $\v{\theta}^{\NE}$. Any other action induces negative reward and CTS yields the efficient schedule. Notice how the payoff of maximal player ($B_3$) changes in the intermediate regime, which leads to her choosing $\theta_m^{\NE}>0$. This results in efficiency loss of CTS. Interestingly, the maximum payoff for all players (and highest efficiency loss for CTS) is attained at the low liquidity regime in Figure \ref{fig:low}, which can result either from a small number of players or small budgets. This outcome is problematic from a market design perspective: it incentivizes players' to misrepresent $B_i$'s. Understanding how selection of both $(\theta_i, B_i)$ impacts CTS outcomes is an interesting direction for future research. However, the conclusions drawn here call for more attention from SOs and regulators in the design of such market mechanisms.

\subsection{Learning equilibria through repeated play}
\label{sec:simulation}
Nash equilibria characterize how the incentives of market participants are oriented. However, the power of said equilibria to predict market outcomes may appear limited in that players are endowed with  intelligence over their opponents' payoff and the system conditions to compute such an equilibrium. In practice, players interact repeatedly exploring the market environment while facing a noisy reward. Motivated to investigate if players can learn equilibria through repeated play, we study the game dynamics where bidders adopt action-value methods (see \cite{sutton}) to update their bids. More precisely, we implement an upper confidence bound (UCB) algorithm for each bidder. In such a setting, each player is agnostic to the presence of other players and the SOs' clearing process, i.e., they endogenize these as part of the environment that yields a random reward. UCB is a popular reinforcement learning algorithm that achieves logarithmic regret per \cite{Auer2010, Lai} in static environments and balances between {exploration} and {exploitation}. In each round (an instance of a CTS market), each player selects the action that has the maximum observed payoff thus far plus some exploration bonus.

The game proceeds as follows. At each round, each bidder chooses $\theta$ from a finite set of actions $\Theta := \{ \theta^1, \ldots,\theta^M\}$. Each bidder maintains a vector $\v{R} \in \Rset^M$ of average rewards  from each action and the number of times $\v{T} \in \Nset^M$ each action is chosen, where $\Nset$ denotes the set of naturals. Here, the reward equals the revenue less the transaction cost from the CTS market. Bidders initialize $\v{R}$ by selecting every action (possible bid from $\Theta$) at least once. Upon bidding $\theta^k \in \Theta$ at a certain round, say she receives the reward $r^k$ from the CTS market. Then, the bidder updates $T^k$ and $R^k$ as
\begin{equation}
T^k \gets T^k + 1, \quad 
R^k \gets R^k + \dfrac{1}{T^k} \left( r^k - R^k \right).
\label{eq:Qlearn}
\end{equation}
Then, the bidder bids the action $\theta^k$, where
\begin{equation}
k =  \underset{j \in \{1,\ldots,M\}}{\text{argmax}} \left\{ R^j + \rho  \sqrt{{\ln \left(\bone^\T \ \v{T} \right)}/{T^j}} \right\}, \label{eq:ucb}
\end{equation}
The parameter $\rho >0$ controls the degree of exploration. The larger the $\rho$, the player is eager to explore actions that have not been tried often enough. The smaller the $\rho$, the player tends to choose an action largely based on the average reward seen thus far.
%
%
\begin{figure}[t!]
	\centering
	\subfloat[Subfigure 1 list of figures text][{Intermediate liquidity}]{
		\includegraphics[width=0.4\textwidth]{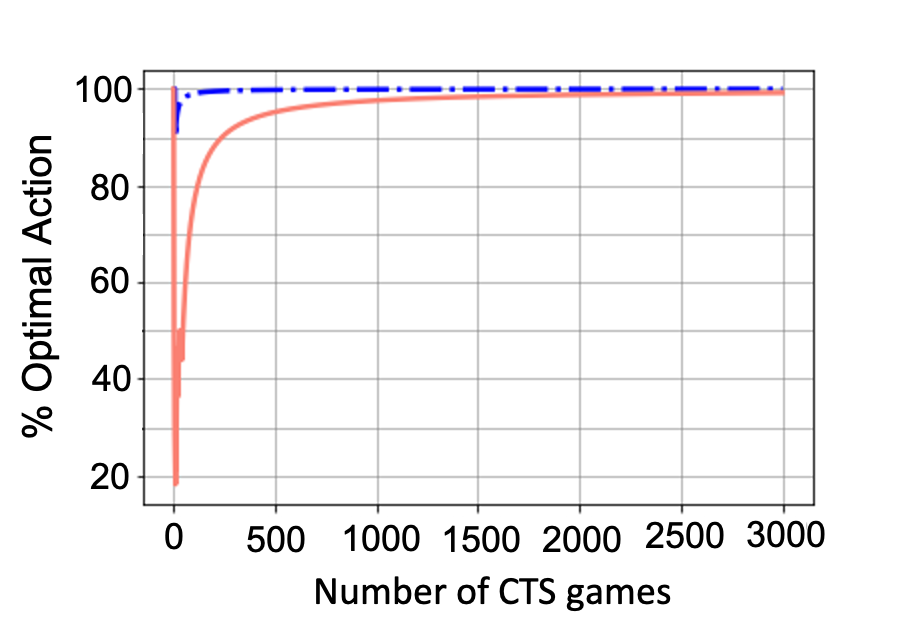}
		\label{fig:illiquid}}
	\subfloat[Subfigure 2 list of figures text][{High liquidity}]{
		\includegraphics[width=0.4\textwidth]{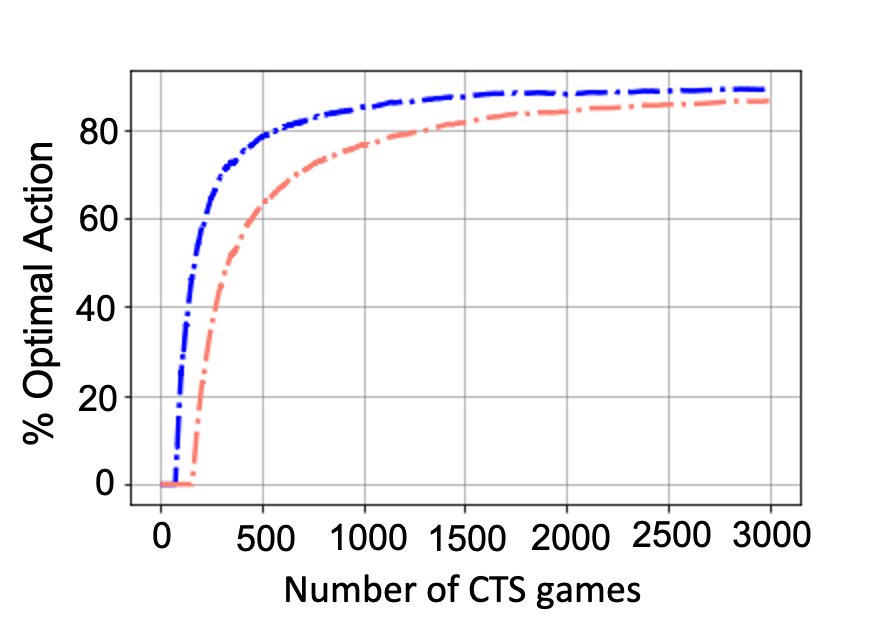}
		\label{fig:liquid}}
	\caption{Plot of cumulative percentage of times the Nash action is chosen across 3000 games for bidders 1 (\protect\redline) and 5 (\protect\blueline). Bidder 5 is marginal for (a) and inframarginal for (b).  After 3000 games, bidders 1-5 respectively select $\v{\theta}^{\NE}$ in (99.9, 92.1, 99.9, 99.6, 99.2)$\%$ games in (a) and (90.1, 99.9, 86.4, 92.4, 88.2)$\%$ games in (b).}
	\label{fig:sim1}
	\centering
	\subfloat[Subfigure 1 list of figures text][Tie-line schedules]{
		\includegraphics[width=0.4\textwidth]{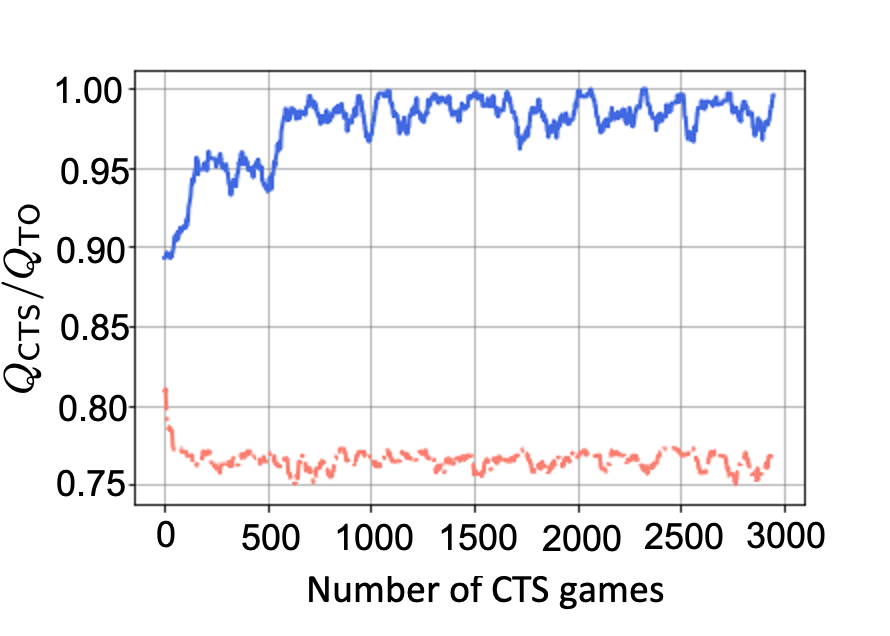}
		\label{fig:qcts}}
	\subfloat[Subfigure 2 list of figures text][Price spread]{
		\includegraphics[width=0.4\textwidth]{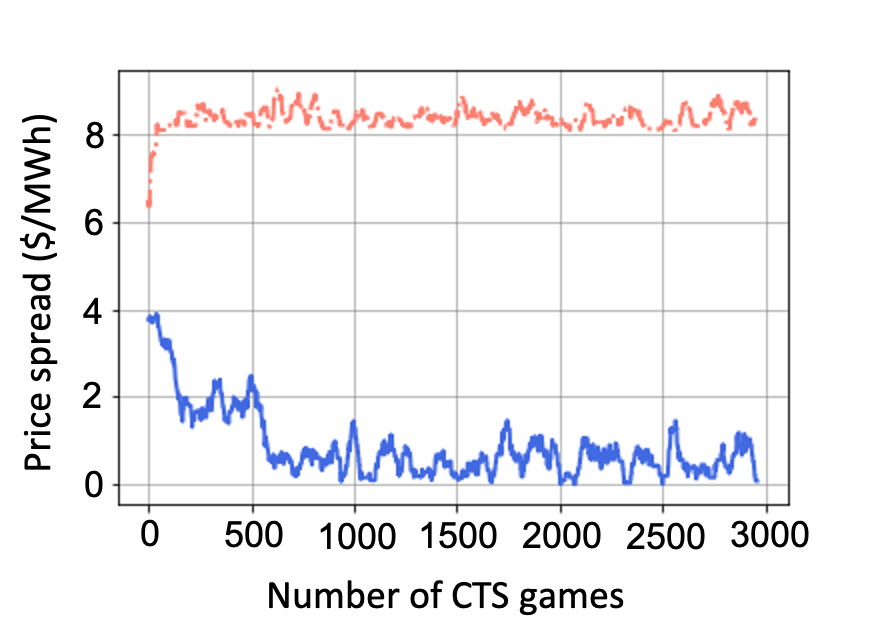}
		\label{fig:spread}}
	\caption{Comparison of tie-line schedules and price spreads for a highly (\protect\myblueline) and intermediately liquid (\protect\blueline) CTS market.}
	\label{fig:sim4}
\end{figure}

We utilize historical CTS data from the NYISO and ISONE markets to compute the affine price spread that yields $Q_{\TO} = 1493$ MW. We consider repeated play of the CTS game with five participants, first with $\v{B} = \left(298, 223, 194, 149, 893\right)$ and then with $\v{B} = \left(596, 522, 640, 373, 893\right)$. The first example corresponds to an intermediate liquidity regime with $\v{\theta}^{\NE} = \left(0, 0, 0, 0, 4882\right)$. The second example belongs to the high liquidity category for which $\v{\theta}^{\NE} = \left(0, 0, 0, 0, 0\right)$. In our simulations, we use $\rho=2$ following \cite[Chapter 2]{sutton}. Each CTS bidder chooses from ten $\theta$'s in $\Theta = \left[0, 6000\right]$ that includes the optimal actions. Figure \ref{fig:sim1} shows percentages of optimal actions selected by bidders in a total of 3000 games for the high and intermediate liquidity regimes.

In the intermediate regime, the pivotal and inframarginal players act in a rather `greedy' fashion, exploiting their optimal action north of $99\%$ of the games. This implies that the observed reward from playing the optimal action is large enough, even as the exploration bonus of other actions increases. Bidder 5 loses her role as the marginal player when the liquidity is high. In this regime, players are slower to discover their optimal actions, although selection percentages are north of $88\%$ of the games. 

Our numerical experiments clearly demonstrate that even in a setting where players know little to nothing about the game setting, they are able to discover and play equilibrium actions (in majority of the games) through repeated play. This experiment lends credence to the conclusions from our equilibrium analysis.
Indeed, $Q_{\CTS}/Q_{\TO}$ in Figure \ref{fig:sim4} remains close to unity and price spreads are below $\$2/$MWh in most games for a highly liquid CTS market. A liquidity reduction of around $40\%$ has palpable effects on market performance, although in aggregate, the players have the capacity to meet $Q_\TO$. In particular, the price spread for intermediate liquidity is more than $\$6/$MWh higher than the highly liquid case and $Q_{\CTS}/Q_{\TO}$ remains well below $80\%$. This experiment highlights how rise of pivotal players exercising market power exploiting the lack of liquidity can impact market performance.

\section{Interactions with Virtual Trading in Energy Markets}
\label{sec:FTR}

CTS performance can be influenced by uneconomic bidding that aims to benefit financial positions of virtual transactions in energy markets. An example of said transactions are up-to-congestion (UTC) virtual bids \cite{UTC}. A UTC is a bid in the day-ahead market to purchase congestion between two nodes within each area. The UTC bid consists of a specified source and sink location together with a price spread that identifies how much the participant is willing to pay for congestion between source and sink. The payoff of a UTC bid depends on the real-time and day-ahead prices at the specified locations.

Bidding behavior in CTS markets impacts CTS outcomes, that in turn affect price movements in both areas. Said price movements influence the return from UTC positions. Thus, bidders with existing UTC portfolios can engage in uneconomic bidding behavior. Here, we utilize our game model to illustrate one such case, where UTC positions negatively impact CTS performance. We remark that price manipulation via uneconomic virtual transactions has emerged as a central policy concern for FERC; several high-profile enforcement cases have ended in multi-million dollar settlements \cite{Ledgerwood}. 

Denote by ${f}^k_{i}$, the UTC megawatt position of CTS bidder $i$ from an internal node $k$ inside area $b$ to the CTS trading location. Let $\Pcal^k_b$ denote the LMP at node $k$ in area $b$. Denote by $\Pcal_b^{k,\da}$ and $\Pcal_{b}^{\da}$ the day-ahead prices at internal node $k$ and CTS trading location, respectively. Then, the payoff of bidder $i$ from her UTC positions is given by
\begin{equation}
\sum_{k}\[\(\Pcal_b - \Pcal_b^k\) - \(\Pcal_b^{\da} - \Pcal_b^{k,\da}\)\] f_i^k,
\end{equation}
where the sum is taken with $k$ ranging over buses within area $b$. The CTS outcome will not affect day-ahead prices, but it does influence real-time prices at other locations inside each area. We have assumed so far that $\Pcal_b - \Pcal_a$ has an affine dependence on $Q$. Assume a similar affine dependence
$$ \Pcal_b(Q) - \Pcal^k_b(Q) = \alpha_{\textrm{in}}^k - \beta_{\textrm{in}}^k Q$$
between the CTS trading location and an internal node $k$ in area $b$. Albeit simplistic, this model is enough to reveal the impact of UTCs on CTS markets. To illustrate the coupling between UTC positions and CTS market, consider the joint payoff from them for bidder $i$ in
\begin{align}
\widetilde{\pi}_{i}(\theta_i, \v{\theta}_{-i})  
= & \underbrace{(\alpha - \beta Q)  B_i - {\theta_i}}_{\text{from CTS}} \notag \\ & \ \ \ \ + \underbrace{\textstyle\sum_{k} ( \alpha^k_{\textrm{in}} - \beta^k_{\textrm{in}} Q)  f_{i}^k - (\Pcal_b^{\da} - \Pcal_b^{k,\da})f_i^k}_{\text{from UTC}},
\label{eq:payoff.total.a}
\end{align}
where $Q$ depends on CTS market clearing with bids $\v{\theta}$ and liquidities $\v{B}$. Formally, call this game $\Gcal_\utc \left( \v{B}, c, \alpha, \beta, \v{f}, \v{\alpha}_{\textrm{in}}, \v{\beta}_{\textrm{in}}\right)$ with payoffs in \eqref{eq:payoff.total.a}. Here, $\v{\alpha}_{\textrm{in}}$, $\v{\beta}_{\textrm{in}}$, $\v{f}$ collect the respective variables across all internal buses. Our next result characterizes the market outcome with UTC positions.

\begin{proposition}
	\label{prop:ftr}
	The game $\Gcal_{\utc}\left({\v{B}}, 0, \alpha, \beta,\v{f}, \v{\alpha}_{\textrm{in}}, \v{\beta}_{\textrm{in}}\right)$ admits a unique Nash equilibrium if $\v{f}$ is elementwise nonnegative, for which the tie-line schedule at the equilibrium is
	\begin{align*}
	Q_{\CTS} &= \begin{cases}
	Q_{\TO}, &\text{if } \bone^\T \ \v{B} - \widetilde{B}_m \geq \alpha / \beta, \\
	\frac{1}{2} \left( Q_{\TO} + \bone^\T \ \v{B} - \widetilde{B}_m\right), & \text{if } \left| \bone^\T \ \v{B} - {\alpha}/{\beta} \right| < \widetilde{B}_m, \\
	\bone^\T \ \v{B}, &\text{otherwise},
	\end{cases}
	\end{align*}
	where $\widetilde{B}_i = B_i + \sum_{k}({\beta_{\textrm{in} }^k}/{\beta}) f_i^k$ for $i=1,\ldots, N$ and $m$ is the only player with maximal $\widetilde{B}_m$.
\end{proposition}
The result reveals that the bidder with maximum combined CTS and UTC position emerges as the pivotal player in this market. Moreover, $\widetilde{B}_m \geq B_m$ dictates that less power is scheduled to flow in the tie-line when bidders have such UTC positions. This results from the incentives of the pivotal player who benefits from higher prices at the importing region $b$'s CTS bus as that yields a higher UTC payoff. In fact, the difference in the tie-line schedules with and without UTC, grows with $\widetilde{B}_m- B_m$ that is directly proportional to the UTC positions. Opposite conclusions can be drawn if we consider players with UTC positions that source at area $b$'s CTS trading node. 

The following example illustrates the shift in market power and scheduling efficiency when participants hold UTCs. Consider the CTS market in Section \ref{sec:simulation} where the fifth bidder is pivotal in the intermediate liquidity regime. At the equilibrium, $Q_{\CTS} = 1176$ MW. 
Assume that the first bidder holds a UTC $f_1 = 800$ MW to an internal bus for which $\alpha_{\textrm{in}} =35.7$ and $\beta_{\textrm{in}} = 0.02$. Then, $\widetilde{\v{B}} = \left[1018, 463, 193, 149, 893\right]$. Notice that bidder one emerges as the new marginal bidder and has incentive to bid in a way that leads to less power being scheduled to flow into area $b$. Indeed, the new tie-line schedule is $Q_{\CTS} = 1113$ MW, $63$ MW less than CTS without UTCs, falling even shorter of $Q_\TO = 1493$ MW.

\section{Impact of Forecast Errors \& Transaction Costs}
\label{sec:conjectures}
Our analysis of the CTS game so far has assumed that players and the SOs have perfect forecasts into the price spread function. In practice, tie-line scheduling takes place with a lead time to power delivery, meaning that there is an inherent uncertainty in the price spread when these markets are convened. To model this uncertainty, assume that the SOs conjecture an affine price spread function 
$$\Pcal_{\SO}(Q) = \alpha_{\SO} - \beta_{\SO} Q$$ 
with $\alpha_{\SO}, \beta_{\SO} > 0$. The SOs use this spread to clear the CTS market as in \eqref{eq:marketProblem}. 
Let the realized price difference be 
$$\mathcal{P}_\star(Q) = \alpha_\star - \beta_\star Q$$ 
with $\alpha_\star, \beta_\star > 0$.
Then, the TO schedule and the optimal tie-line schedule, respectively, are given by 
$$ Q_\TO = \alpha_\SO/ \beta_\SO \quad \text{and} \quad Q_\star = \alpha_\star/\beta_\star. $$
Modeling the uncertainty explicitly at the time of scheduling reveals that $Q_\TO$ may not equal $Q_\star$, the ex-post optimal tie-line schedule. Our interest lies in analyzing if strategic behavior of bidders in the CTS market can correct the errors in SOs' forecasts. Do bidders draw the outcome closer to $Q_\star$ than $Q_\TO$ or do they drive it further away as a result of their strategic interaction? We answer this question through a game-theoretic study. We also derive insights into how non-zero transaction fees ($c >0$) affect these conclusions.

To isolate the impacts of uncertainty and transaction fees, we analyze the game under a simpler setting where the bidders are homogenous, each with liquidity $B>0$ and conjectured price spread $\mathcal{P}(Q) = \alpha - \beta Q$ with $\alpha, \beta > 0$. 
Notice that bidders' conjectured optimal schedule $\alpha/\beta$ may be different from both $Q_\TO$ and $Q_\star$. We assume here that players share a common belief that the market operates at an intermediate liquidity where the aggregate liquidity $NB$ is close to her conjectured optimal tie-line schedule $\alpha/\beta$, i.e., 
\begin{align}
N B = \alpha/\beta + \Ocal(1/N).
\label{eq:assn_liquidity}
\end{align}
Under such an assumption, bidder $i$ conjectures the market price from bidding $\v{\theta}$ with liquidities $\v{B} = B\bone$ to be
\begin{align*}
p\left(\v{\theta}, B\bone \right) 
&= \dfrac{1}{2}\left( \mathcal{P}(NB) + \sqrt{\mathcal{P}^2(NB) + 4 \beta \bone^\T \ \v{\theta}} \right)\\
&= \sqrt{\beta \bone^\T \ \v{\theta}} + \Ocal(1/N),
\end{align*}
which yields the following perceived payoff for bidder $i$.
\small
\begin{align}
\pi_{i}(\theta_i, \v{\theta}_{-i}) 
&= p(\v{\theta}, \v{B}) B -\theta_i - c\left(B - \dfrac{\theta_i}{p\left(\v{\theta}, B\bone\right)}\right) \notag \\
&\approx \sqrt{\beta \bone^\T \ \v{\theta}} B -\theta_i - c\left(B - \dfrac{\theta_i}{\sqrt{\beta \bone^\T \ \v{\theta}}}\right).
\label{eq:payoff.heterogeneous}
\end{align}
\normalsize
Call the CTS game with conjectured price spreads $\Gcal_{\textrm{conj}}(B, c, \alpha, \beta, \alpha_\SO, \beta_\SO)$, where $\alpha, \beta$ satisfy \eqref{eq:assn_liquidity} and the payoffs are given by \eqref{eq:payoff.heterogeneous}. Assuming that all players offer based on an equilibrium profile for this game, the SOs then solve the CTS flow allocation problem in \eqref{eq:marketProblem} with $\Pcal_\SO$ to ultimately compute the tie-line schedule. Our next result characterizes both a (symmetric) equilibrium profile and the resulting tie-line schedule.
\begin{proposition}
	\label{prop:heterogeneous}
	The CTS game $\Gcal_{\textrm{conj}}(B, c, \alpha, \beta, \alpha_\SO, \beta_\SO)$ admits a unique symmetric Nash equilibrium given by
	$\theta^\NE_i = \frac{\gamma^2}{4 N \beta}$
	for $i=1,\ldots,N$, for which the tie-line schedule at equilibrium is 
	\begin{align}
	Q_{\CTS} = \dfrac{1}{2}\left[Q_{\TO} + N B - \sqrt{  \left(Q_\TO - NB\right)^2 + \dfrac{\gamma^2}{\beta \beta_{\SO}} } \right],\notag \label{eq: QCTS.het}
	\end{align}
	where $\gamma :=  c(2-1/N) + \beta B$. 
\end{proposition}
Our proof leverages the result in \cite[Theorem 3]{symmetric} and the analysis of first-order conditions for a symmetric equilibrium of the game. Notice that players bid solely based on their own conjectures. The tie-line schedule, however, depends on the conjectures of both the bidders and the SOs. This result will allow us to study the effect of price spread forecasts and transaction costs on the scheduling efficiency in the sequel.

\begin{figure}[!h]
	\centering
	\vspace{-0.1in}
	\includegraphics[width=0.5\textwidth]{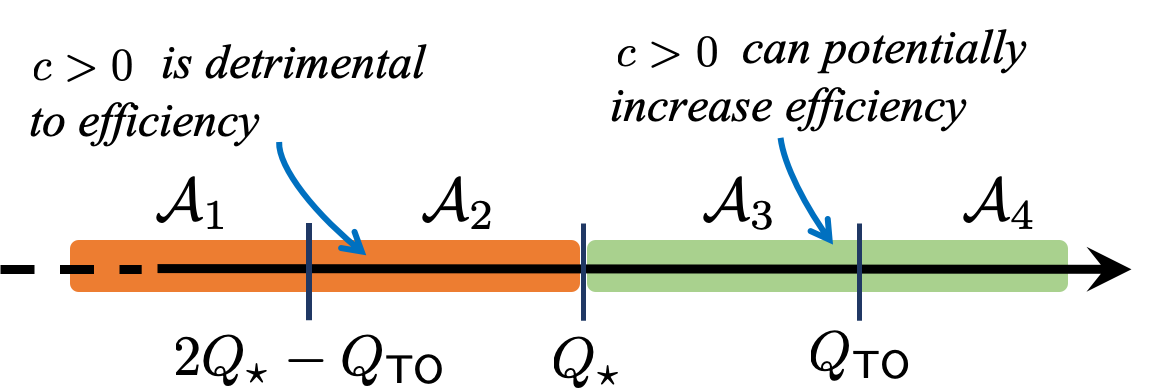}
	\caption{Ability of market participants to correct SO's forecast error depends on liquidity and transactions costs.}
	\label{fig:cost}
\end{figure}

The lack of knowledge of $Q_\star$ by the SOs and market participants prompts us to investigate whether CTS can yield a more efficient schedule than the pure SO-driven TO. Proposition \ref{prop:heterogeneous} implies $Q_{\CTS}\leq Q_\TO$, meaning that CTS cannot yield a more efficient schedule than TO if $Q_\TO < Q_\star$. Hence, CTS can only outperform TO when the SOs' forecast overestimates $Q_{\TO}$. In this regime, Figure \ref{mechanisms} yields that $Q_{\CTS}$ is always closer to $Q_\star$ when $Q_\star \leq Q_{\TO}/2$. Outside of this setting, the outcome of CTS depends on the liquidity and conjectures of players. 
Specifically, if $NB \in \mathcal{A}_1 \cup \mathcal{A}_2$, defined in Figure \ref{fig:cost}, $Q_{\CTS}$ is closer to $Q_\star$ than $Q_\TO$, if 
\begin{equation}
\dfrac{\gamma^2}{\beta \beta_{\SO}} \leq 8 \left( Q_{\TO} - Q_{\star} \right) \left(Q_{\TO} - 2Q_\star + NB\right).
\label{eq:condition}
\end{equation}
Such a premise appears to run counter to the intuition that TO is optimal. This situation can only arise under uncertainty where SOs make serious forecast errors in the expected price spread. Surprisingly, forecast errors are not that rare, according to \cite{nyiso}, where the error in SOs' point forecast for the price spread between NYISO and ISO-NE averaged $\$2.42/$MWh. Notice how, in this liquidity regime, the presence of transaction fees makes it harder to satisfy \eqref{eq:condition}. This is intuitively correct since transaction fees drive the tie-line schedule toward smaller values, as established in Proposition \ref{prop:heterogeneous}. 
When $NB \in \mathcal{A}_3 \cup \mathcal{A}_4$, liquidity is sufficiently high and the presence of costs might improve scheduling efficiency since players bid higher prices to counter costs. Overall, players ability to correct SOs' forecast is somewhat limited and relies on many qualifications, indicating that the SOs forecasts and systematic bias plays a vital role in scheduling efficiency. Moving bid submittal and clearing timelines closer to power delivery should improve the efficiency of CTS.
\begin{figure}[h!]
	\centering
	\includegraphics[width=0.5\textwidth]{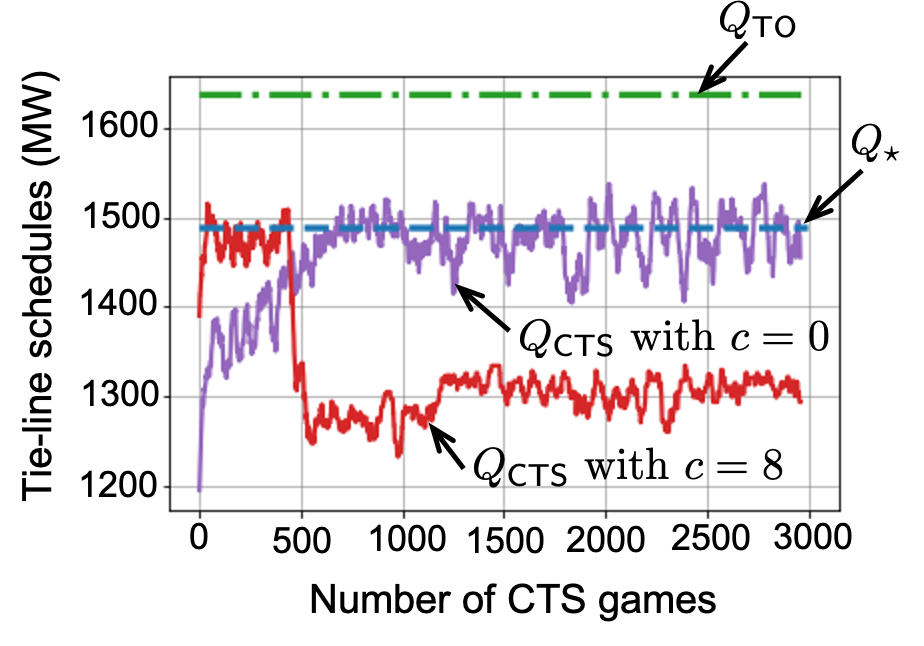}
	\caption{The trajectory of CTS schedules cleared against SO's forecasted prices with $10\%$ error with $c=0$ and $c=\$8/$MWh.}
	\label{fig:conjecture}
\end{figure}

Proposition \ref{prop:heterogeneous} suggests that incentives of CTS bidders are aligned in a way that allows them to correct SOs' forecast errors in some settings. 
Can players learn such equilibria through repeated play. We employ the learning framework in Section \ref{sec:simulation}, where players have their bids cleared against $(\alpha_{\SO}, \beta_{\SO})$ that are perturbed from  $(\alpha_\star, \beta_\star)$ learned from historical data. That is, in every round, bidders receive reward from the ex-post price spread described by $\Pcal_\star$. The trajectory of tie-line schedules in Figure \ref{fig:conjecture} with $c=0$ reveals that bidding behavior of players results in CTS schedules consistently closer to the ex-post optimal than TO. 
Despite the SO's persistent forecast error, bidders `correct' the tie-line schedule to an extent by seeking actions that maximize their observed reward.

The relation in \eqref{eq:condition} reveals that presence of nonzero transaction fees $c$ make it more difficult for CTS market to drive the outcome closer to the ex-post optimal as $\gamma$ increases with $c$. Bidders reacting to observed rewards with $c = \$8$/MWh in Figure \ref{fig:conjecture} yield a CTS schedule farther from $Q_\star$, seeking actions that yield higher prices but smaller schedules. This result corroborates our theoretical finding that transaction fees impede bidders' ability to correct SOs' forecast errors.
\begin{figure}[h!]
	\centering
	\subfloat[Subfigure 1 list of figures text][]{
		\includegraphics[width=0.4\textwidth]{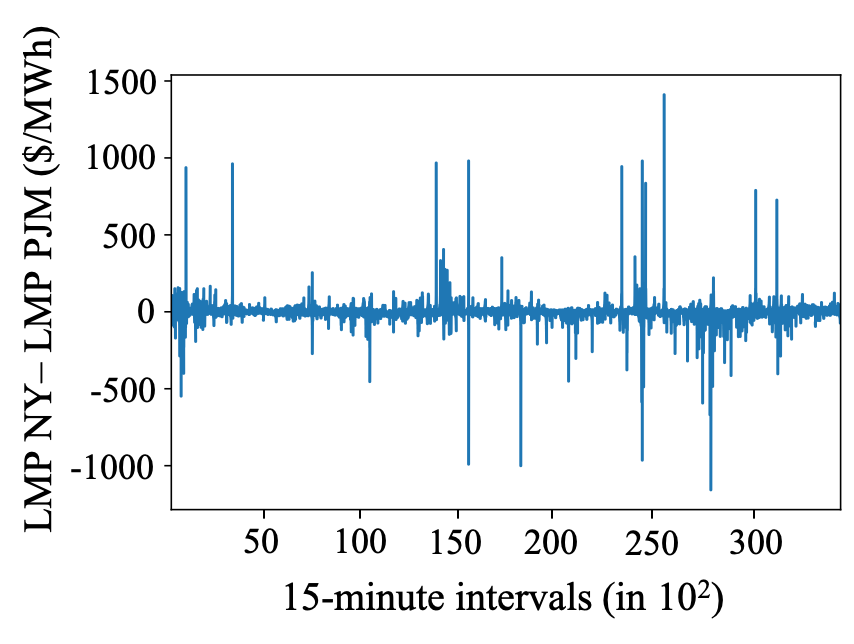}
		\label{fig:SpreadNYPJM}}
	\subfloat[Subfigure 2 list of figures text][]{
		\includegraphics[width=0.4\textwidth]{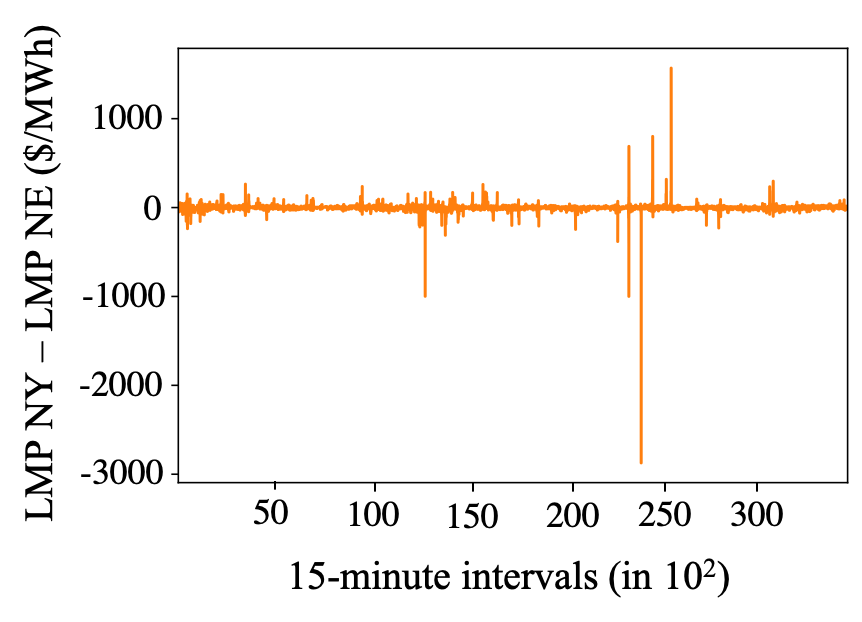}
		\label{fig:SpreadTimeSeries}}
	\caption{Plot (a) depicts the time series of spread between NYISO and PJM proxy buses in 2018 (absolute mean = 8.92 \$/MWh, std. deviation = 22.11 \$/MWh). Plot (b) shows the same between NYISO and ISO-NE for the same year (mean = 0.44, absolute mean = 5.59  \$/MWh, std. dev. = 18.14 \$/MWh).}
\end{figure}

Notice that equilibrium bid grows with $c$, per Proposition \ref{prop:heterogeneous}. With $c > 0$, bidders are reluctant to offer their entire liquidity. A similar result can be shown under more general settings of Theorem \ref{th:general}. This may prevent the price spread from converging to zero, even if the market is liquid. And, transaction fees make it less attractive for CTS bidders overall, hurting long-term liquidity of the CTS market.  Figure \ref{fig:SpreadNYPJM} indicates that the price spread in the CTS market between NYISO and PJM exhibits longer excursions from zero and higher volatility compared to that between NYISO and ISO-NE, depicted in Figure \ref{fig:SpreadTimeSeries}. The average absolute spread between NYISO and PJM is approximately \$3.3/MWh higher than that between NYISO and ISO-NE. We surmise that transaction fees between NYISO and PJM and the lack thereof between NYISO and ISO-NE are largely responsible for this difference.



\bibliographystyle{abbrvnat}
\bibliography{myrefs}


\appendix

\section{Proof of Theorem \ref{th:general}.}

We break the proof into two parts--for $c >0$ and $c = 0$. We argue that $\Gcal(\v{B}, c)$ is a concave game with a compact strategy set $\v{\Theta}$ in each case. Then, the rest follows from Rosen's result by \cite[Theorem 1]{rosen}. First, we establish that the objective function of \eqref{eq:marketProblem} is strictly concave for all $\bone^\T \ \v{\theta} > 0$. Denote by $\psi$, the objective of \eqref{eq:marketProblem} and let $\v{H}$ denote its Hessian. Then, the entries of $\v{H}$ are given by
\begin{align*}
H_{ii} = \dfrac{\partial^2 \psi}{\partial x_i^2} = \dfrac{\partial \Pcal}{\partial x_i} - \dfrac{\theta_i}{ (B_i - x_i)^2 }, \ \
H_{ij} = \dfrac{\partial ^2 \psi}{\partial x_i \partial x_j} = \dfrac{\partial \Pcal}{ \partial x_j}
\end{align*}
for $i,j =1, \ldots, N, j \neq i$.
Since $\dfrac{\partial \Pcal}{\partial x_i} = \Pcal'$, we have 
\begin{equation}
\v{y}^\T \ \v{H} \v{y} = - \sum_{i=1}^{N} \dfrac{\theta_i}{ (B_i - x_i)^2  }y_i^2 + \dfrac{\partial \Pcal}{\partial Q} \( \sum_{i=1}^{N} y_i\)^2 < 0
\end{equation}
for any $\v{y} \neq \v{0}$. Thus, $\psi$ is strictly concave in $\v{x}$. 

\underline{Case with $c>0$:} Notice that in this case $\bone^\T \ \v{\theta} > 0$. Since, the objective of \eqref{eq:marketProblem} is strictly concave, it follows that if a solution to the maximization problem exists, then it is unique. The first-order optimality conditions of \eqref{eq:marketProblem} yield that such an optimal allocation $\v{x}^\star$ must satisfy
\begin{equation}
\Pcal(\bone^\T \ \v{x}^\star) - \dfrac{\theta_i}{B_i - x_i^\star} = 0, \ \forall i.
\label{eq:x.star}
\end{equation}
Summing the above relation over $i$, we get
\begin{equation}
\Pcal(Q_{\CTS}) = \dfrac{\bone^\T \ \v{\theta}}{ \bone^\T \ \v{B} - Q_{\CTS}}.
\label{eq:Qstar}
\end{equation}
Since $\Pcal$ is strictly decreasing with $\Pcal(0) > 0$, the strictly increasing function of $Q_{\CTS}$ that grows to $\infty$ at $\bone^\T \ \v{B}$ in the RHS of \eqref{eq:Qstar} must intersect $\Pcal$ at a unique point. Thus, $Q_{\CTS}$ is uniquely defined for each $\v{\theta}$, and so is $\v{x}^\star$ identified by
\begin{equation}
x_i^\star = B_i - \dfrac{\theta_i}{\bone^\T \ \v{\theta}} \( \bone^\T \ \v{B} - Q_{\CTS} \).
\label{eq:x.star.solution}
\end{equation}
To establish that $\Gcal(\v{B}, c)$ is a concave game, we now establish that the payoffs $\pi(\theta_i, \v{\theta}_{-i})$ in \eqref{eq:payoff.general} are continuous is $\v{\theta}$ and concave in $\theta_i$. Notice that the unique optimal allocation $\v{x}^\star$ is continuous in $\v{\theta}$, owing to Berge's maximum theorem (see \cite{Ok}), implying the same for $Q_{\CTS}$. In turn, that proves the continuity of $\pi_i$ in $\v{\theta}$. Next, we prove that $\pi_i$ is concave in $\theta_i$, by showing that $\Pcal(Q_{\CTS})$ is concave and $x_i^\star$ is convex in $\theta_i$.

First, we show that $\Pcal(Q_{\CTS})$ is concave. Notice that
\begin{align*}
\frac{\partial^2}{\partial \theta_i^2}\Pcal(Q_{\CTS}) 
= 
\Pcal^{''}(Q_{\CTS}) \left(\frac{\partial Q_\CTS}{\partial \theta_i} \right)^2 + \Pcal'(Q_{\CTS}) \frac{\partial^2 Q_\CTS}{\partial \theta_i^2}
\end{align*}
Since $\Pcal$ is concave and strictly decreasing, it suffices to show that $Q_\CTS$ is convex in $\theta_i$ to conclude that $\frac{\partial^2}{\partial \theta_i^2}\Pcal(Q_{\CTS}) \leq 0$ and hence,  $\Pcal(Q_{\CTS})$ is concave in $\theta_i$.  

To prove the convexity of $Q_{\CTS}$ in $\theta_i$, rewrite \eqref{eq:Qstar} as $g(Q_{\CTS}) = \bone^\T \ \v{\theta}$, where
\begin{equation}
g(Q_{\CTS}) :=  (\bone^\T \ \v{B} - Q_{\CTS}) \mathcal{P}(Q_{\CTS}).
\label{eq:solutionQ}
\end{equation}
Now, $g(0) >0$ and $g$ is a continuous and strictly decreasing function of its argument. Also, $g$ is convex because
\begin{equation}
g''(Q_{\CTS}) = \Pcal ''(Q_{\CTS}) (\bone^\T \ \v{B} - Q_{\CTS}) - 2 \mathcal{P}'(Q_{\CTS}) \geq 0, \label{eq:condition.apx}
\end{equation}
where the inequality follows from the requirement in \eqref{eq:existence}, the strictly decreasing and concave nature of $\Pcal$, and the non-negativity of $\bone^\T \ \v{B} - Q_{\CTS}$. 
These derivatives exist, owing to the implicit function theorem (see \cite{Mas}). Then, $Q_{\CTS}$ is the inverse of a decreasing convex function, and is therefore decreasing convex itself in $\bone^\T \ \v{\theta}$, and therefore in $\theta_i$. This completes the proof of the concavity of $\Pcal(Q_{\CTS})$. 

Next, we show that $x_i^\star$ is convex in $\theta_i$. 
From \eqref{eq:x.star.solution}, 
we get 
\begin{align}
\dfrac{\partial x_i^{\star}}{\partial \theta_i} &= - \dfrac{\bone^\T \ \v{\theta}_{-i}}{\left(\bone^\T \ \v{\theta}\right)^2}(\bone^\T \ \v{B} - Q_{\CTS}) + \dfrac{\theta_i}{\bone^\T \ \v{\theta}} \dfrac{\partial Q_{\CTS}}{\partial \theta_i},\label{eq:x.prime}
\\
\dfrac{\partial^2 x_i^\star}{\partial \theta_i^2} &= 2 \dfrac{\bone^\T \ \v{\theta}_{-i}}{\left(\bone^\T \ \v{\theta}\right)^3}(\bone^\T \ \v{B} - Q_{\CTS}) + 2 \dfrac{\bone^\T \ \v{\theta}_{-i}}{\left(\bone^\T \ \v{\theta}\right)^2} \dfrac{\partial Q_{\CTS}}{\partial\theta_i} \notag \\ & \ \ \ \ \ \ \ \ \ \ \ + \dfrac{\theta_i}{\bone^\T \ \v{\theta}} \dfrac{\partial ^2Q_{\CTS}} {\partial \theta_i^2}. 
\label{eq:second.derivative.x}
\end{align}
Again, the implicit function theorem guarantees that these derivatives exist for $\v{\theta}$ away from the origin. 
The last term in \eqref{eq:second.derivative.x} is non-negative by convexity of $Q$. Therefore, we require the sum of the remaining terms to be non-negative or
$$ \dfrac{\bone^\T \ \v{B} - Q_{\CTS}}{\bone^\T \ \v{\theta}} \geq - \dfrac{\partial Q_{\CTS}(\v{\theta}; \v{B})}{\partial \theta_i}.$$ 

From \eqref{eq:solutionQ} we have
\begin{align*}
\dfrac{\bone^\T \ \v{B} - Q_{\CTS}}{\bone^\T \ \v{\theta}} & = \dfrac{1}{\mathcal{P}(Q_{\CTS})}  \notag \\ 
& \geq \dfrac{1}{\mathcal{P}(Q_{\CTS})-\mathcal{P}'(Q_{\CTS})(\bone^\T \ \v{B} - Q_{\CTS})}  \notag \\
& = - \dfrac{\partial Q_{\CTS}(\v{\theta}, \v{B})}{\partial\theta_i},
\end{align*}
where the inequality follows from the fact that $\mathcal{P}'(Q_{\CTS})(\bone^\T \ \v{B} - Q_{\CTS}) < 0$. Finally, notice that $Q_{\CTS} \rightarrow 0$ and $\Pcal \rightarrow \Pcal(0)$ as $\theta_i \rightarrow \infty$. It follows that $x_i^\star \rightarrow 0$ and the payoff in \eqref{eq:payoff.general} goes to negative infinity as $\theta_i$ grows unbounded. Therefore, there exists $\theta_i^{\textrm{max}}$ such that for $\theta_i \geq \theta_i^{\textrm{max}}$, $\pi_i$ becomes negative. As such, player $i$ would never choose $\theta_i \geq \theta_i^{\textrm{max}}$ at an equilibrium and we can restrict the strategy space of each CTS player to the compact interval $[cB_i , \theta_i^{\textrm{max}}]$. Applying Rosen's result \cite[Theorem 1]{rosen} we establish existence of a Nash equilibrium for $c>0$.

\underline{Case with $c=0$:} The payoff $\pi_i$ is continuous in $\v{\theta}$ and concave in $\theta_i$ for all $\bone^\T \ \v{\theta} > {0}$. We extend the same to $\bone^\T \ \v{\theta} = {0}$ with $c=0$. With zero costs, we have
\begin{equation}
\pi_i(\theta_i, \v{\theta}_{-i}) = \Pcal(Q_{\CTS}(\v{\theta}; \v{B})  ) B_i - \theta_i. \label{eq:payoff.zero.c}
\end{equation}
It suffices to argue that $Q_\CTS$ is continuous at $\v{\theta} = \v{0}$. Recall that for $\bone^\T \ \v{\theta} > 0$, $Q_{\CTS}$ is given by
\begin{equation}
\bone^\T \ \v{\theta} = \( \bone^\T \ \v{B} - Q_{\CTS} \) \Pcal(Q_{\CTS}). \label{eq:auxiliary}
\end{equation}

First, assume that $\bone^\T \ \v{B} < Q_{\TO}$. Then, $\Pcal (Q_{\CTS}) \geq \Pcal(\bone^\T \ \v{B}) > 0$ since $\Pcal$ is strictly decreasing.
Consider a sequence $\v{\theta}_k \rightarrow \v{0}$ as $k \rightarrow \infty$. Then, the LHS of \eqref{eq:auxiliary} vanishes. Therefore, the RHS must vanish as well. Since $\Pcal$ does not vanish, we must have $Q_{\CTS}(\v{\theta}_k) \rightarrow \bone^\T \ \v{B} = Q_{\CTS}(\v{0})$ as required. Now, consider the situation where $\bone^\T \ \v{B} > Q_{\TO}$. In this case, $Q_{\CTS} \leq Q_{\TO} < \bone^\T \ \v{B}$. Consider the sequence $\v{\theta}_k \rightarrow \v{0}$ as $k \rightarrow \infty$. As the LHS of \eqref{eq:auxiliary} vanishes, $\Pcal$ must vanish in the RHS. Therefore, $\Pcal(Q_{\CTS}(\v{\theta}_k)) \rightarrow 0$ or $Q_{\CTS}(\v{\theta}_k) \rightarrow \Pcal^{-1}(0) = Q_{\TO} = Q_{\CTS}(\v{0})$, as required for the case with $\bone^\T \ \v{B} > Q_{\TO}$. The case with $\bone^\T \ \v{B} = Q_{\TO}$ is trivially satisfied by the same line of arguments. Hence, $Q_{\CTS}(\v{\theta}; \v{B})$ is continuous in $\v{\theta}$ at the origin. Moreover, with the same line of arguments as in the case with $c>0$, we can restrict the strategy space of each player to the compact interval $[0, \theta_i^{\textrm{max}}]$.
This completes the proof.

\section{Proof of Proposition \ref{prop:Nash}}
Existence of the Nash equilibrium follows from Theorem \ref{th:general}.
Solving \eqref{eq:marketProblem} we find that $Q_{\CTS}(\v{\theta},\v{B} )$ is given by
\begin{equation}
Q_{\CTS}(\v{\theta}, \v{B}) = \dfrac{\alpha + \beta \bone^\T \ \v{B}}{2 \beta} - \dfrac{1}{2 \beta}\left\lbrack\mathcal{P}^2\left(\bone^\T \ \v{B}\right) + 4\beta\bone^\T \ \v{\theta}\right\rbrack^{1/2}. \label{eq:CTS.schedule.affine}
\end{equation}
The payoff for player $i$ is given by
\begin{align}
\pi_{i}(\theta_i, \v{\theta}_{-i}) & = \left(\alpha - \beta Q_{\CTS}(\v{\theta}, \v{B})\right) B_i - \theta_i \notag \\
&=\dfrac{B_i}{2}\left(\mathcal{P}(\bone^\T \ \v{B}) + \left\lbrack\mathcal{P}^2(\bone^\T \ \v{B}) + 4 \beta \bone^\T \ \v{\theta}\right\rbrack^{1/2}\right) - \theta_i. 
\end{align}
The payoff is continuous in $\v{\theta}_{-i}$ and strictly concave in $\theta_i$. Note that $\pi_{i}(\theta_i, \v{\theta}_{-i})$ becomes negative for $\theta_i > \beta B_i^2$. Hence, we restrict our attention to for a Nash equilibrium in the compact interval $\left[0, \beta B_i^2\right]$.
A bid profile $\v{\theta}^{\NE} = \left( \theta_1^{\NE}, \ldots, \theta_N^{\NE}\right)$ is a Nash equilibrium if and only if
\begin{subequations}
	\begin{equation}
	\left. \dfrac{\partial \pi_{i}(\theta_i, \v{\theta}_{-i})}{\partial \theta_i}\right \vert_{\v{\theta}^\NE} \leq 0, ~\text{ if } 0 \leq \theta_i^{\NE} < \beta B_i^2
	\end{equation}
	\begin{equation}
	\left. \dfrac{\partial \pi_{i}(\theta_i, \v{\theta}_{-i})}{\partial \theta_i}\right \vert_{\v{\theta}^\NE} \geq 0, ~\text{ if } 0 < \theta_i^{\NE} \leq \beta B_i^2, 
	\end{equation}\label{eq:EC}
\end{subequations}
where the above derivative is given by
\begin{equation}
\dfrac{\partial \pi_{i}(\theta_i, \v{\theta}_{-i})}{\partial \theta_i} = \dfrac{\beta B_i}{\left\lbrack \mathcal{P}^2(\bone^\T \ \v{B}) + 4 \beta \bone^\T \ \v{\theta}  \right\rbrack^{1/2}  } - 1. \label{eq:payoff.Der}
\end{equation}
From \eqref{eq:payoff.Der} we deduce that the payoff derivative cannot vanish for more than one player. Moreover, no player would bid $\theta_i^{\NE} = \beta B_i^2$ since that yields negative payoff and each player profitably deviates by infinitesimally decreasing $\theta_i$. From the previous discussion and the following observation 
\begin{equation}
\dfrac{\partial \pi_m (\theta_m, \v{\theta}_{-m})}{\partial \theta_m} > \dfrac{\partial \pi_i (\theta_i, \v{\theta}_{-i})}{\partial \theta_i}, ~ i \neq m \label{eq:monotonicity}
\end{equation}
we conclude that $\v{\theta}_{-m}^{\NE} = \mathbf{0}$. In search for positive $\theta_m >0$ we find that
\begin{itemize}
	\item If $\left|\bone^\T \ \v{B} - \alpha/\beta\right| < B_m$, then 
	\begin{equation}
	\theta_m^{\NE} = \dfrac{\beta^2 B_m^2 - \mathcal{P}^2(\bone^\T \ \v{B})}{4 \beta} >0. \label{eq:thetaNash}
	\end{equation}
	\item Otherwise, $\theta_m^{\NE} = 0$ since \eqref{eq:thetaNash} yields a negative value.
\end{itemize}
To prove the bounds on $\eta_{\CTS}(\v{B})$ first note that the social welfare attains its maximum at $Q = Q_{\TO}$ with
\begin{equation}
\Wcal(Q_{\TO}) = \dfrac{\alpha^2}{2 \beta}.
\end{equation}
Hence, in the high liquidity regime, i.e., $\bone^\T \ \v{B} - B_m \geq \alpha /\beta$, $Q_{\CTS} = Q_{\TO}$ and $\eta_{\CTS}(\v{B}) = 1$.
In the intermediate regime, the social welfare at $Q_{\CTS}$ is
\begin{align}
\Wcal(Q_{\CTS}) = \frac{\alpha}{2}\left(\frac{\alpha}{\beta} + \bone^\T \ \v{B}_{-m}\right) - \frac{\beta}{8} \left( \frac{\alpha}{\beta} + \bone^\T \ \v{B}_{-m} \right)^2 &= \frac{3}{4} \left( \dfrac{\alpha^2}{2 \beta}\right) + \dfrac{\bone^\T \ \v{B}_{-m}}{4} \left( \alpha - \dfrac{1}{2}\beta \bone^\T \ \v{B}_{-m}\right)\notag \\& \ > \dfrac{3}{4} \Wcal(Q_{\TO}).
\end{align}

Finally, in the low liquidity regime, i.e., $\bone^\T \ \v{B} + B_m \leq \alpha /\beta$, we have
\begin{equation}
\dfrac{\Wcal(Q_{\CTS})}{\Wcal(Q_{\TO})} = \dfrac{1}{\alpha^2} \left(2 \beta (\bone^\T \ \v{B}) \left(\alpha - \dfrac{\beta}{2} \bone^\T \ \v{B}\right) \right) = \dfrac{2 \beta \bone^\T \ \v{B} }{\alpha} - \dfrac{ \beta^2 (\bone^\T \ \v{B})^2}{\alpha^2} = 2x - x^2. \label{eq:efficiency1}
\end{equation}

\section{Proof of Proposition \ref{prop:ftr}}

It is easy to verify that \eqref{eq:payoff.total.a} is concave in $\theta_i$ for fixed $\v{\theta}_{-i}$ and $\v{f}$ nonnegative. Moreover, $Q$ is strictly decreasing in $\theta_i$ and as $\theta_i$ grows large the price spreads approach the limiting values $\alpha$ and $\alpha^k$. Hence, in \eqref{eq:payoff.total.a} the first two terms converge to constant values with the affine term approaching negative infinity as $\theta_i$ grows unbounded. Therefore, there exists $\theta_i^{\textrm{max}}$ such that \eqref{eq:payoff.total.a} becomes negative for $\theta_i \geq \theta_i^{\textrm{max}}$. As scuh, we restrict our attention for a Nash equilibrium within the compact interval $\left[0, \theta_i^{\textrm{max}}\right]$. Existence of a Nash equilibrium for $\Gcal_{\utc}\left(\widetilde{\v{B}}, 0, \alpha, \beta, \v{\alpha}^k, \v{\beta}^k\right)$ is established by invoking \cite[Theorem 1]{rosen}.
A bid profile $\v{\theta}^{\NE} = \left( \theta_1^{\NE}, \ldots, \theta_N^{\NE}\right)$ is a Nash equilibrium if and only if \eqref{eq:EC} are satisfied where ${\pi}_{i}$ is replaced with $\widetilde{\pi}_{i}$ and $\beta B_i^2$ with $\theta_i^{\textrm{max}}$.
The payoff derivative is given by
\begin{align}
\dfrac{\partial \widetilde{\pi}_{i}(\theta_i, \v{\theta}_{-i})}{\partial \theta_i} &= \dfrac{\beta\left(B_i + \sum_{k}^{} \frac{\beta_{\textrm{in}}^k}{\beta} f_i^k \right)}{ \left\lbrack \mathcal{P}^2(\bone^\T \ \v{B}) + 4 \beta \bone^\T \ \v{\theta}  \right\rbrack^{1/2}  }  - 1 \notag \\ &= \dfrac{\beta \widetilde{B}_i}{ \left\lbrack \mathcal{P}^2(\bone^\T \ \v{B}) + 4 \beta \bone^\T \ \v{\theta}  \right\rbrack^{1/2}  }  - 1. 
\label{eq:payoff.Der.ftr}
\end{align}
The rest of proof is similar to that of Proposition \ref{prop:Nash}.

\section{Proof of Proposition \ref{prop:heterogeneous}}
We are in search for a symmetric equilibrium for $\Gcal_{\textrm{conj}}(B, c, \alpha, \beta, \alpha_\SO, \beta_\SO)$. From first order conditions we find that the payoff's derivative is given by
\begin{equation}
\dfrac{\partial \pi_{i}(\theta_i, \v{\theta}_{-i})}{\partial \theta_i} = \dfrac{\beta B}{2 p\left(\v{\theta}, B\bone \right) } -1 + c\left\lbrack\dfrac{1}{p\left(\v{\theta}, B\bone \right) } - \dfrac{\theta_i}{2 p\left(\v{\theta}, B\bone \right)  \bone^\T \ \v{\theta}} \right\rbrack,\label{eq:dpdtheta}
\end{equation}
where $p\left(\v{\theta}, B\bone \right) = \sqrt{\beta \bone^\T \ \v{\theta}}$.
For $\theta_i^{\NE}>0$ we require \eqref{eq:dpdtheta} to vanish, yielding the following
\begin{equation}
\dfrac{\theta_i^{\NE}}{\bone^\T \ \v{\theta} ^{\NE}} = \dfrac{\beta B}{c} + 2 - \dfrac{2}{c} \sqrt{\beta \bone^\T \ \v{\theta}}.\label{eq:theta.Nash}
\end{equation}
Summing \eqref{eq:theta.Nash} over $i$'s we find
\begin{equation}
\sqrt{\bone^\T \ \v{\theta}^{\NE}} = \dfrac{1}{N\sqrt{\beta}} \left(\dfrac{N B \beta}{2} + \dfrac{c}{2}(2N-1) \right) >0. \label{eq:theta.sum}
\end{equation}
From \eqref{eq:theta.sum} and \eqref{eq:theta.Nash} we find that
\begin{equation}
\theta_i^{\NE} = \dfrac{1}{4 N \beta} \left( \beta B + c(2 - \frac{1}{N})\right)^2, \label{eq:bored}
\end{equation}
which is strictly positive. The solution of \eqref{eq:marketProblem} with $\Pcal_{\SO}$ yields the CTS schedule
\begin{equation}
Q_{\CTS} = \dfrac{1}{2}\left(Q_{\TO} + NB \right) - \dfrac{1}{2 \beta_{\SO}} \sqrt{ \left(\alpha_{\SO} - \beta_{\SO}NB \right)^2 + 4\beta_{\SO} \bone^\T \ \v{\theta} }.\label{eq:verybored}
\end{equation}
Substituting \eqref{eq:bored} in \eqref{eq:verybored} we obtain the expression in Proposition \ref{prop:heterogeneous}.

\end{document}